\documentclass[12pt]{article}
\usepackage{a4wide}
\usepackage{bbm}
\usepackage{amssymb}

\newcommand{\0}{\hbox{\kern2.5pt\vrule height 7.5pt\kern-2.5pt 0}}
\newcommand{\1}{\mathbbm{1}}
\newcommand{\C}{\mathbbm{C}}
\newcommand{\E}{\mathbbm{E}}
\renewcommand{\H}{\mathbbm{H}}
\newcommand{\M}{\mathbbm{M}}
\newcommand{\N}{\mathbbm{N}}
\renewcommand{\P}{\mathbbm{P}}
\newcommand{\R}{\mathbbm{R}}
\newcommand{\Z}{\mathbbm{Z}}
\newcommand{\Lor}{\mathop{\rm Lor}\nolimits}
\newcommand{\lOR}{\mathop{\rm lor}\nolimits}
\newcommand{\Bor}{\mathop{\rm Bor}\nolimits}
\newcommand{\bor}{\mathop{\rm bor}\nolimits}
\newcommand{\Tor}{\mathop{\rm Tor}\nolimits}
\newcommand{\tor}{\mathop{\rm tor}\nolimits}

\newcommand{\car}{\mathop{\rm car}\nolimits}

\newcommand{\GL}{\mathop{\rm GL}\nolimits}
\newcommand{\gL}{\mathop{\rm gl}\nolimits}
\newcommand{\SL}{\mathop{\rm SL}\nolimits}
\newcommand{\sL}{\mathop{\rm sl}\nolimits}
\newcommand{\SO}{\mathop{\rm SO}\nolimits}
\newcommand{\so}{\mathop{\rm so}\nolimits}
\newcommand{\SU}{\mathop{\rm SU}\nolimits}
\newcommand{\su}{\mathop{\rm su}\nolimits}
\newcommand{\spec}{\mathop{\rm spec}\nolimits}

\newcommand{\tr}{\mathop{\rm tr}\nolimits}

\newcommand{\diag}{\mathop{\rm diag}\nolimits}
\newcommand{\eps}{\varepsilon}
\newcommand{\ID}{\hbox{$I\kern-3pt D$}}
\newcommand{\IID}{\hbox{$I\kern-3pt I\kern-3pt D$}}
\newcommand{\rot}{\mathop{\rm rot}\nolimits}
\newcommand{\ad}{\mathop{\rm ad}\nolimits}
\newcommand{\kest}{\mathop{\rm span}\nolimits}
\newcommand{\sol}{\mathop{\rm sol}\nolimits}
\newcommand{\Sol}{\mathop{\rm Sol}\nolimits}
\newcommand{\longvec}{\raise7pt\hbox to0pt{$\,-\kern-7pt\longrightarrow$\hss}}

\newcommand{\po}{\raise7pt\hbox to0pt{$\scriptstyle\,\circ$\hss}p}
\newcommand{\Lambdao}{\raise7pt%
  \hbox to0pt{\kern1.8pt$\scriptstyle\circ$\hss}\Lambda}
\newcommand{\omegao}{\raise5pt\hbox to0pt{$\scriptstyle\,\circ$\hss}\omega}
\newcommand{\Beps}{B^{(\eps)}}
\newcommand{\beps}{b^{(\eps)}}
\newcommand{\Teps}{T^{(\eps)}}
\newcommand{\teps}{t^{(\eps)}}
\newcommand{\ueps}{u^{(\eps)}}

\begin{document}

\thispagestyle{empty} 
\begin{flushright}
MITP/16-055
\end{flushright}

\begin{center}
{\Large\bf Mass, zero mass and \ldots\ nophysics}\\[1.0cm]
{\large R.~Saar$^1$ and S.~Groote$^{1,2}$}\\[0.3cm]
$^1$ Loodus- ja T\"appisteaduste valdkond, F\"u\"usika Instituut,\\[.2cm]
  Tartu \"Ulikool, W.~Ostwaldi 1, 50411 Tartu, Estonia\\[7pt]
$^2$ PRISMA Cluster of Excellence, Institut f\"ur Physik,\\[.2cm]
Johannes-Gutenberg-Universit\"at,
Staudinger Weg 7, 55099 Mainz, Germany
\end{center}

\vspace{0.2cm}
\begin{abstract}\noindent
In this paper we demonstrate that massless particles cannot be considered
as the limiting case of massive particles. Instead, the usual symmetry
structure based on semisimple groups like $U(1)$, $SU(2)$ and $SU(3)$ has to
be replaced by less usual solvable groups like the minimal nonabelian group
$\sol_2$. Starting from the proper orthochronous Lorentz group $\Lor_{1,3}$ we
extend Wigner's little group by an additional generator, obtaining the maximal
solvable or Borel subgroup $\Bor_{1,3}$ which is equivalent to the Kronecker
sum of two copies of $\sol_2$, telling something about the helicity of
particle and antiparticle states. 
\end{abstract}

\newpage

\section{Introduction}
In his paper ``Sur la dynamique de l'electron'' from July 1905, Henri
Poincar\'e formulates the Principle of Relativity, introduces the concepts of
Lorentz transformation and Lorentz group, and postulates the covariance of
laws of Nature under Lorentz transformations~\cite{Poincare:1905}. In 1939,
Eug\'ene Wigner analysed the unitary representations of the inhomogeneous
Lorentz group~\cite{Wigner:1939cj,Wigner:1962}. However, there are also
nonunitary representations. As Wigner pointed out, the irreducible
representations of the Lorentz group are important because they define the
types of particles. The Lorentz symmetry is expressed by introducing the
commutative diagram for the covariant wave functions $\psi^C$,
\begin{equation}\label{eq01}
\begin{tabular}{rlcr}
$\psi^C:$&$\E_{1,3}\ni p$&$\longrightarrow$&$\psi^C(p)$\\
$\llap{$U(\Lambda)$}\downarrow$&$\downarrow\rlap{$\Lambda$}$&&
  $\downarrow\rlap{$T(\Lambda)$}\quad$\\
$U(\Lambda)\psi^C:$&$\Lambda x\ni\Lambda p$&$\longrightarrow$
&$T(\Lambda)\psi^C(p)$\\
\end{tabular}
\end{equation}
implying $(U(\Lambda)\psi^C)(\Lambda p)=T(\Lambda)\psi^C(p)$. Here
$\Lor_{1,3}\ni\Lambda\to T(\Lambda)$ is the finite dimensional nonunitary
representation.

Let the vectors $|p,\lambda\rangle$ with four-momentum $p$ and independent
parameter $\lambda$ form a complete orthogonal basis of the irreducible
representation of the Poincar\'e group. In this basis, $U(\Lambda)$ is always
expressed by~\cite{Ohnuki:1976,Weinberg:1995}
\[U(\Lambda)=Q\left(W(\Lambda,p)\right)P(\Lambda),\]
where $Q$ is diagonal with respect to $\lambda$, $P$ is diagonal with
respect to $p$, and $W(\Lambda,p)$ is the Thomas--Wigner rotation.

The transformation law for Wigner's wave function $\psi^W$ reads
\begin{equation}\label{eq02}
(U(\Lambda)\psi^W)(p,\lambda)=\sum_{\lambda'}Q_{\lambda\lambda'}
  \left(W(\Lambda,\Lambda^{-1}p)\psi^W(\Lambda^{-1}p,\lambda')\right).
\end{equation}
Here $Q(W(\Lambda,p))$ is a representation of Wigner's little group as a
subrepresentation of some Lorentz group representation $T(\Lambda)$. The
explicit relation between $\psi^C$ and $\psi^W$ is given in
Refs.~\cite{Shaw:1964zz,Niederer:1974ps,Fronsdal:1959zz,Halpern:1965zza,%
Mackey:1969vt,Candlin:1965,Pursey:1965zz,Feldman:1966,Dragon:2016ohn}.

The great importance of Wigner's theory is that the classification of
particles according to their Lorentz transformation properties is entirely
determined by the representation of the little group as the subrepresentation
of the representation of
$\Lor_{1,3}=\SO_{1,3}^0$~\cite{Sternberg:1994,Kim:1986wv}.

The most important cases are
\[\po=(1,\underbrace{0,0,0}_{\llap{$\textstyle{\rm Lg}(\po)\ =\qquad$}
  \textstyle\SO_3}), \underbrace{(1,\underbrace{0,0,1}_{\textstyle E(2)})
  }_{\textstyle\Bor_{1,3}}\]

In the title we use the word ``nophysics''. This word can be taken as an
abbreviation, namely
\[\mbox{nophysics}=\mbox{new old physics}\]
where ``old'' stands for Pauli's massless neutrino hypothesis, ``new'' for
the solvability as symmetry applicable to massless particles. Another more
detailed explanation is found in the Conclusions of this paper.

\section{The little group}
The little group of the four-momentum $\po$ or the stabiliser of $\po$ is the
maximal closed subgroup of $\Lor_{1,3}$ defined as
\begin{equation}\label{eq03}
\mathop{\rm Lg}\po=\{\Lambdao\in\Lor_{1,3}:\ \Lambdao\po=\po\}.
\end{equation}
The orbit of $\po$ is a subspace in $\E_{1,3}$,
\[\mathop{\rm Orb}\po=\{\Lambda\po:\ \Lambda\in\Lor_{1,3}\}=\Lor_{1,3}\po,\]
given as the bijection $\mathop{\rm Orb}\po=\Lor_{1,3}/\mathop{\rm Lg}\po$.
For all $p\in\mathop{\rm Orb}\po$ the Thomas--Wigner rotation is given by
\begin{equation}
W(\Lambda,p)=L^{-1}_{\Lambda p}\Lambda L^{\phantom{-1}}_p,
\end{equation}
where $L_p\in\Lor_{1,3}$ is the representative of $p\in\mathop{\rm Orb}\po$.

The characteristic feature of the massive case is that the fixed vector $\po$
can be chosen to be $\po=(m,\vec 0)$, for which
\[\mathop{\rm Lg}\po=\SO_3\buildrel{\rm locally}\over=\SU_2\]
($SU_2$ is the universal covering group of $SO_3$). Since $SU_2$ is compact
and simply connected, all its finite-dimensional irreducible representations
are single-valued, unitary and parametrised by the eigenvalue $s$ of the
Casimir operator which can take on non-negative half-integer values
\[s=0,\frac12,1,\frac32,\ldots\]
From the local equivalence, the equivalence of the Lie algebras can be
derived, i.e.\
\[\so_3=\mathop{\rm ad}\su_2=\mathop{\rm lg}\po.\]
It is important that the little group $SO_3$ is the maximal compact simple
subgroup of $\Lor_{1,3}$ while $\su_2$ (i.e.\ $\mathop{\rm lg}\po$) is the
simplest semisimple Lie algebra, $\dim_\R\su_2=3$.

For every irreducible unitary representation of the little group
$\mathop{\rm Lg}\po=\SO_3$ one can derive a corresponding induced
representation of the Poincar\'e group,
${\cal P}_{1,3}={\cal T}_{1,3}\rtimes\Lor_{1,3}$. The irreducible
representations of ${\cal P}_{1,3}$ are characterised by the pairs $(m,s)$,
where the mass $m$ is real and positive and the spin takes on the values
$s=0,\frac12,1,\ldots$ The states within each irreducible representation are
labelled by $\xi=-s,-s+1,\ldots,s$ which means that massive particles of spin
$s$ have $2s+1$ degrees of freedom.

In the massless case $m=0$ the representative vector $\po$ may be taken to be
\begin{equation}\label{eq05}
\po=(\omega_p,0,0,\omega_p),\quad \omega=|\vec p\,|.
\end{equation}
To construct the little group, one has to solve the defining equation
\begin{equation}\label{eq06}
\Lambdao\po=\po.
\end{equation}
The result is the Euclidean group~\cite{Wigner:1939cj,Wigner:1962,Ohnuki:1976,%
Weinberg:1995,Shaw:1964zz,Niederer:1974ps,Fronsdal:1959zz,Halpern:1965zza,%
Mackey:1969vt,Candlin:1965,Pursey:1965zz,Feldman:1966,Dragon:2016ohn,%
Sternberg:1994,Kim:1986wv}
\[\mathop{\rm Lg}\po=E(2)=\mathop{\rm ISO}(2)={\cal T}_2\rtimes\SO(2)\]
for which the double covering group is given by
\[\overline E(2)={\cal T}_2\rtimes U(1)\]
where ${\cal T}_2$ is the Abelian two-dimensional group of translations. Thus,
the little group is solvable and non-compact, and the restrictions of the
finite-dimensional representations of $\Lor_{1,3}$ to $E(2)$ are in general
non-unitary. In fact, the only unitary, irreducible representation of $E(2)$
are one-dimensional, i.e.\ degenerate representations, since the subgroup
${\cal T}_2$ of translations has to be realised
trivially~\cite{Ohnuki:1976,Weinberg:1995},
\begin{equation}\label{eq07}
E(2)\rightarrow E(2)/{\cal T}_2=\SO(2).
\end{equation}
The requirement that the representations are at most double-valued implies
that only the representations $\SO(2)\to U^{(j)}(\SO(2))$,
$j=0,\pm\frac12,\pm1,\ldots$ are allowed. This one-dimensional (internal)
freedom of massless particles is usually called the helicity. Since all the
unitary representations on the orbits $p^2=0$ are induced by the
non-faithful one-dimensional representation of the little group $E(2)$,
massless particles are characterised by a discrete helicity
\begin{equation}\label{eq08}
\lambda=0,\pm\frac12,\pm 1,\ldots
\end{equation}
Notice that if the parity is included, the helicity takes on two values,
$\lambda$ and $-\lambda$. For example, the two states $\lambda=\pm1$ are
then referred to as left-handed ($\lambda=-1$) and right-handed ($\lambda=+1$)
photons.

\subsection{The Borel subgroup}
Due to the unitarity of representations of the little group
$\mathop{\rm Lg}\po=E(2)$, zero-mass particles have only a single value for
the helicity (if the parity is not taken into account). Suppose, the most
general determined, relativistically invariant, first order single particle
equation is of the form
\begin{equation}\label{eq09}
\left(\beta^\mu\partial_\mu+\rho\right)\psi(x)=0.
\end{equation}
Then there is a simple criterion by D.~Kwoh under which this equation will
have zero-mass solutions~\cite{Wightman:1973},

\vspace{12pt}\noindent
{\bf Kwoh's lemma:} A necessary condition that Eq.~(\ref{eq09}) has a
zero-mass solution is that
\begin{equation}\label{eq10}
\det(-i\beta^\mu p_\mu+\lambda\rho)=0
\end{equation}
for all real $\lambda$ and all light-like $p$, i.e.\ all $p$ such that $p^2=0$.

If Eq.~(\ref{eq09}) is a defining equation for a single massless particle,
Kwoh's lemma states the gauge invariance $p\to\lambda p$ as a very special
property of the theory. Therefore, it seems to be reasonable (at least
mathematically) to include this gauge transformation into the little group,
i.e.\ instead of Eq.~(\ref{eq06}) take
\begin{equation}\label{eq11}
\Lambdao\po=\lambda\po
\end{equation}
as defining equation for the little group, where $\po=(\eps,0,0,1)$ with
$\eps=\pm 1$, $\lambda>0$ and $\Lambdao\in\Lor_{1,3}$. If
$\Lambda=\exp(-\frac12\omega^{\mu\nu}e_{\mu\nu})$, Eq.~(\ref{eq11}) yields
$\omega^\mu{}_\nu\po^\nu=\delta\lambda\po^\mu $ or more explicitly
\[\omega_{03}=\delta\lambda\eps,\qquad\eps\omega_{01}-\omega_{13}=0,\qquad
\eps\omega_{02}-\omega_{23}=0\]
(cf.\ Sec.~\ref{lor13gen}). Notice that in case of $E(2)$ as little group one
has $\delta\lambda=0$. The solution of Eq.~(\ref{eq11}) reads
\begin{equation}\label{eq12}
\Lambdao=\Beps(\vec\xi;\lambda,\omega)
  =\pmatrix{\frac12(\frac1\lambda\xi^2+\lambda+\frac1\lambda)
  &\eps\vec\xi^{\,T}\rot\omega
  &\frac\eps2(-\frac1\lambda\xi^2+\lambda-\frac1\lambda)\cr
  \frac\eps\lambda\vec\xi&\rot\omega&-\frac1\lambda\vec\xi\cr
  \frac\eps2(\frac1\lambda\xi^2+\lambda-\frac1\lambda)
  &\vec\xi^{\,T}\rot\omega
  &\frac12(-\frac1\lambda\xi^2+\lambda+\frac1\lambda)\cr},
\end{equation}
where
\[\vec\xi=\pmatrix{\xi_1\cr\xi_2\cr},\quad
\rot\omega=\pmatrix{\cos\omega&-\sin\omega\cr \sin\omega&\cos\omega\cr},\quad
\xi^2=\xi_1^2+\xi_2^2.\]
If one now defines
\begin{eqnarray}\label{eq13}
\Beps_\lambda&=&\Beps(0;\lambda,0)
  \ =\ \pmatrix{\frac12(\lambda+\frac1\lambda)&\vec 0^{\,T}
  &\frac\eps2(\lambda-\frac1\lambda)\cr \vec 0&\1_2&\vec 0\cr
  \frac\eps2(\lambda-\frac1\lambda)&\vec 0^{\,T}
  &\frac12(\lambda+\frac1\lambda)\cr},\nonumber\\
R_\omega&=&\Beps(0;1,\omega)
  \ =\ \pmatrix{1&\vec 0^{\,T}&0\cr \vec 0&\rot\omega&\vec 0\cr
  0&\vec 0^{\,T}&1\cr},\nonumber\\
T^{(\eps)}_\xi&=&\Beps(\vec\xi;1,0)
  \ =\ \pmatrix{1+\frac12\xi^2&\eps\vec\xi^{\,T}&-\frac\eps2\xi^2\cr
  \eps\vec\xi&\1_2&-\vec\xi\cr
  \frac\eps2\xi^2&\vec\xi^{\,T}&1-\frac12\xi^2\cr},
\end{eqnarray}
the general transformation~(\ref{eq12}) can be written as
\[\Beps(\vec\xi;\lambda,\omega)=T^{(\eps)}_\xi \Beps_\lambda
  R_\omega.\]
One easily obtains the multiplication table
\begin{eqnarray}\label{eq14}
\Beps_\mu\Beps_\lambda&=&\Beps_{\mu\lambda}\ =\ \Beps_\lambda\Beps_\mu,
  \nonumber\\[7pt]
R_\phi R_\omega&=&R_{\phi+\omega}\ =\ R_\omega R_\phi,\nonumber\\[7pt]
\Teps_\eta\Teps_\xi&=&\Teps_{\eta+\xi}\ =\ \Teps_\xi\Teps_\eta,\nonumber\\[7pt]
\Beps_\lambda\Teps_\xi&=&\Teps_{\lambda\xi}\Beps_\lambda,\nonumber\\[7pt]
R_\omega\Teps_\xi&=&\Teps_{\rot\omega\xi}R_\omega,\nonumber\\[7pt]
\Beps(\vec\xi;\lambda,\omega)\Beps(\vec\eta;\mu,\varphi)
  &=&\Beps(\vec\xi+\lambda\rot\omega\vec\eta;\lambda\mu,\omega+\varphi),
  \nonumber\\[7pt]
(\Beps(\vec\xi;\lambda,\omega))^{-1}
  &=&\Beps(-\frac1\lambda\rot(-\omega)\vec\xi;\frac1\lambda,-\omega).
\end{eqnarray}
From the multiplication table~(\ref{eq14}) it follows that the transformations
$\Beps(\vec\xi;\lambda,\omega)$ form a group
$\Bor_{1,3}^{(\eps)}\subset\Lor_{1,3}$
with non-compact parameter space
\[\{\vec\xi\in\R_2,\ 0\le\omega\le\pi,\ \lambda>0\}.\]
It can easily be shown that the derived series of commutators ${\cal D}$ for
$\Bor_{1,3}^{(\eps)}$ ends in the identity ${\rm id}$. In fact,
\[{\cal D}^2(\Bor_{1,3}^{(\eps)})=\{{\rm id}\}.\]
Actually, $\Bor_{1,3}^{(\eps)}$ is a maximal, solvable and non-compact
subgroup of $\Lor_{1,3}$, i.e.\ the non-compact Borel subgroup of
$\Lor_{1,3}$. Moreover, one obtains the Borel decomposition as the semidirect
product
\begin{equation}\label{eq15}
{\cal B}^{(\eps)}\equiv\Bor_{1,3}^{(\eps)}
  ={\cal T}_2^{(\eps)}\rtimes\Tor_{1,3}^{(\eps)}.
\end{equation}
The set ${\cal T}_2^{(\eps)}=\mathop{\rm Gen}\{\Teps_\xi:\ \vec\xi\in\R_2\}$
of unipotent elements of $\Bor_{1,3}^{(\eps)}$ is a closed nilpotent subgroup
of $\Bor_{1,3}^{(\eps)}$. It contains the subgroup
${\cal D}(\Bor_{1,3}^{(\eps)})=(\Bor_{1,3}^{(\eps)},\Bor_{1,3}^{(\eps)})$
generated by the commutators and is normal in $\Bor_{1,3}^{(\eps)}$.
$\Tor_{1,3}^{(\eps)}=\Bor_{1,3}^{(\eps)}/{\cal T}_2^{(\eps)}
=\mathop{\rm Gen}\{\Beps_\lambda,\ R_\omega:\ \Lambda>0;\ 0\le\omega\le\pi\}$
is the maximal torus in $\Bor_{1,3}^{(\eps)}$ (and in $\Lor_{1,3}$) with
dimension $\dim(\Bor_{1,3}^{(\eps)}/{\cal T}_2^{(\eps)})$, generated by the
semisimple elements $\Beps_\lambda$ and $R_\omega$. By the Lie--Kolchin
theorem as it is written down later, $\Bor_{1,3}^{(\eps)}$ is upper
triangular~\cite{Borel,TauYu,GoodmanWallach,Serre:1965}.

At this point a remark of S.~Weinberg is in order~\cite{Weinberg:1988fs}:
``For the case of zero mass there are interesting complications. The little
group as Wigner pointed out is a non-semi-simple group, and one must make
special remarks about its invariant Abelian subalgebra.'' Indeed,
\[\mathop{\rm Lg}\po\sim{\rm Abelian}_2\rtimes{\rm Abelian}_2.\]
is the semidirect product~(\ref{eq15}) of two two-parametric Abelian groups.

\subsection{Jordan factorisation}
The Jordan factorisation of $M\in\GL_4(\R)$ into a semisimple and an unipotent
component is given by
\[M=M_uM_s.\]
Since $\Bor_{1,3}$ is solvable, according to the Lie--Kolchin theorem a basis
can be chosen with respect to which $B\in\Bor_{1,3}$ can be put into a
triangular form
\[B=\pmatrix{*&*&*&*\cr 0&*&*&*\cr 0&0&*&*\cr 0&0&0&*\cr}
  =\pmatrix{1&*&*&*\cr 0&1&*&*\cr 0&0&1&*\cr 0&0&0&1\cr}
  \pmatrix{\lambda_0&0&0&0\cr\ 0&\lambda_1&0&0\cr 0&0&\lambda_2&0\cr
  0&0&0&\lambda_3\cr}=B_uB_s\]
where $B_u$ is unipotent (i.e.\ all eigenvalues of $B_u$ are $1$) and
$B_s=\mathop{\rm diag}(\lambda_0,\lambda_1,\lambda_2,\lambda_3)$ is
semisimple. The eigenvalues of $B$ and $B_s$ are identical. In this form, the
Jordan decompositon is given by
\[\Bor_{1,3}\subset U_4(\R)\rtimes D_4(\R)\equiv T_4(\R),\]
where $U_4(\R)$ is the group of upper triangular unipotent matrices and
$D_4(\R)$ is the group of invertible diagonal matrices. If $B$ is solvable,
then~\cite{Jauch:1959dbc}
\[B=\pmatrix{\lambda_0&*&*&*\cr 0&\lambda_1&*&*\cr 0&0&\lambda_2&*\cr
  0&0&0&\lambda_3\cr}\quad\Rightarrow\quad
  B_u=B\pmatrix{\lambda_0&0&0&0\cr\ 0&\lambda_1&0&0\cr 0&0&\lambda_2&0\cr
  0&0&0&\lambda_3\cr}^{-1}\]
and $\det B=1$. From Refs.~\cite{Borel,TauYu,GoodmanWallach,Serre:1965,Bump}
we take the following set of well-known theorems, lemmas and propositions.

\vspace{12pt}\noindent
{\bf Theorem:} Let G be a connected linear algebraic group. Then $G$ contains
a Borel subgroup ${\cal B}$, and all other Borel subgroups of $G$ are
conjugate to ${\cal B}$. The homogeneous space $G/{\cal B}$ is a projective
variety (\cite{GoodmanWallach}, p.~524).

\vspace{12pt}\noindent
{\bf Lie--Kolchin theorem:} If $\pi:\ G\to\GL(V)$ is a linear representation
of a connected solvable group, then $\pi(G)$ leaves a flag in $V$ invariant,
i.e.\ $\pi(G)$ can be put in triangular form (see e.g.\ \cite{TauYu}, p.~406).

\vspace{12pt}\noindent
{\bf Borel Fixpoint Theorem:} Let $S$ be a connected, solvable group that
acts algebraically on a projective variety $X$. Then there exists a point
$x\in X$ such that (\cite{Borel}, p.~137)
\[\Lambda x=x\quad\mbox{for all }\Lambda\in S.\]

Therefore, Eq.~(\ref{eq11}) is reasonable, as the semi-invariant of
$\Bor_{1,3}$ in $\E_{1,3}$ is the non-zero vector $\po$ spanning the
${\cal B}$-stable line in $\E_{1,3}$ ($x=\R\po$).

\vspace{12pt}\noindent
{\bf Lemma:} Let $V$ be a $\C$-vector space of dimension $n>0$ and $S$ a
connected, solvable subgroup of $\GL(V)$. Then there exists a vector
$v\in V\backslash\{0\}$ such that (\cite{TauYu}, p.~407)
\[Sv=\C v.\]
Let ${\cal B}_s$ denote the set of semisimple elements of ${\cal B}$ and
${\cal B}_u$ the set of unipotent elements.

\vspace{12pt}\noindent
{\bf Proposition:} If ${\cal B}$ is connected and solvable, then the set
${\cal B}_u$ is a closed, connected and nilpotent subgroup of ${\cal B}$,
containing ${\cal D}(G)$ and, therefore, is normal in ${\cal B}$
(\cite{TauYu} p.~407).

From this it follows that the set ${\cal B}_s$ is not a closed subgroup of
${\cal B}$ because if it would be a subgroup, ${\cal B}$ would be nilpotent.

\vspace{12pt}\noindent
{\bf Proposition:} Let ${\cal B}$ be connected and solvable and let
$\mathfrak b\equiv{\cal L}({\cal B})$ be its Lie algebra. Then the set
${\cal L}({\cal B}_u)$ is the set of nilpotent elements of $\mathfrak g$
(i.e.\ ${\rm ad}_{\mathfrak g}u$ is nilpotent for $u\in{\cal L}({\cal B}_u)$
(\cite{TauYu}, p.~409).

\vspace{12pt}
It is important that for solvable ${\cal B}$ the set ${\cal B}_u$ of unipotent
elements of ${\cal B}$ is a connected, closed, normal and nilpotent subgroup
of ${\cal B}$, ${\cal B}/{\cal B}_u$ is a torus, and
${\cal D}({\cal B})\subset{\cal B}_u$.

\vspace{12pt}\noindent
{\bf Theorem (Borel):} Let ${\cal B}$ be a connected, solvable, linear
algebraic group. If $\Tor$ is a maximal torus of ${\cal B}$, then
\[{\cal B}={\cal B}_u\rtimes\Tor.\]
Otherwise, there exists such a torus $\Tor\subset{\cal B}$ such that
\[{\cal B}=\mathop{\rm Rad}\nolimits_u({\cal B})\rtimes\Tor,\]
where $\mathop{\rm Rad}_u({\cal B})$ is the unipotent radical of ${\cal B}$,
leading to the factorisation~(\ref{eq15}) (this theorem can be found in the
original work Ref.~\cite{Borel}, p.~137 as well as in Ref.~\cite{TauYu},
p.~410).

The generators of the Borel subgroup in the representation~(\ref{eq13}) are
defined by
\begin{equation}\label{eq16}
\beps_\mu=\frac{\partial\Beps(\omega_\mu)}{\partial\omega_\mu}\Big|_{\omegao},
\end{equation}
\addtocounter{equation}{-1}
where $\omega_0=\lambda$, $\omega_1=\xi_1$, $\omega_2=\xi_2$, $\omega_3=\omega$
and $\omegao=(1,0,0,0)$. This yields the Lie algebra $\bor_{1,3}^{(\eps)}$ as
basis for the underlying vector space to be generated by
\begin{eqnarray}
\beps_0&=&\frac{\partial\Beps_\lambda}{\partial\lambda}\Big|_{\lambda=1}
  \ =\ \eps e_{03},\nonumber\\
\beps_1&=&\frac{\partial T_\xi}{\partial\xi_1}\Big|_{\vec\xi=\vec 0}
  \ =\ \eps e_{01}+e_{31},\nonumber\\
\beps_2&=&\frac{\partial T_\xi}{\partial\xi_2}\Big|_{\vec\xi=\vec 0}
  \ =\ \eps e_{02}+e_{32},\nonumber\\
\beps_3&=&\frac{\partial R_\omega}{\partial\omega}\Big|_{\omega=0}
  \ =\ e_{21}.
\end{eqnarray}
The commutator relations are ($a,b\in\{1,2\}$)~\cite{Finkelstein:1955zz,%
Winternitz:1965,Patera:1975,Lastaria:1999,Zhang:2012hh}
\begin{eqnarray}\label{eq17}
[\beps_0,\beps_a]=\beps_a,&&[\beps_0,\beps_3]=0,\nonumber\\\
[\beps_3,\beps_a]=-\epsilon_{3ab}\beps_b,&&[\beps_a,\beps_b]=0.
\end{eqnarray}
The algebra $\bor_{1,3}$ is solvable because
\[[{\cal D}(\bor_{1,3}^{(\eps)}),{\cal D}(\bor_{1,3}^{(\eps)})]
  ={\cal D}^2(\bor_{1,3}^{(\eps)})=\{0\}\]
and maximal in $\Lor_{1,3}$, i.e.\ the Borel algebra of $\lOR_{1,3}$. Moreover,
\begin{equation}\label{eq18}
\bor_{1,3}^{(\eps)}=\mathfrak t_2^{(\eps)}\rtimes\tor_{1,3}^{(\eps)},
\end{equation}
where the vector space underlying $\mathfrak t_2^{(\eps)}$ is
$\vec{\mathfrak t}_2^{\,(\eps)}=\kest_\R\{\beps_1,\beps_2\}$ and that of
$\tor_{1,3}=\car_{1,3}$ is
$\vec{\mathstrut\tor}_{1,3}^{(\eps)}=\kest_\R\{\beps_0,\beps_3\}$
($\car_{1,3}$ is the Cartan subalgebra of $\lOR_{1,3}$). Therefore, one can
conclude that in the massless case the (enlarged) little algebra
$\bor_{1,3}^{(\eps)}$ is a maximal, non-compact and solvable Lie subalgebra of
$\lOR_{1,3}$, i.e.\ the Borel subalgebra, and is the semidirect sum of two
abelian algebras $\mathfrak t_2^{(\eps)}$ and $\tor_{1,3}^{(\eps)}$. Notice
that there exists no Casimir operator.

In the general case~\cite{Borel,TauYu,GoodmanWallach,Serre:1965}, if
${\cal B}$ is solvable, its Lie algebra $\mathfrak b={\cal L}({\cal B})$ is
solvable. Since ${\cal B}_u$ is normal in ${\cal B}$, its Lie algebra
$\mathfrak n={\cal L}({\cal B}_u)$ is an ideal of $\mathfrak b$ and
$\mathfrak n$ is the set of nilpotent elements of $\mathfrak b$. Moreover,
since ${\cal D}({\cal B})\subset{\cal B}_u$ one has
$[\mathfrak b,\mathfrak b]\subset\mathfrak n$.

\vspace{12pt}\noindent
{\bf Theorem:} There exists a Lie subalgebra $\mathfrak a$ of $\mathfrak b$
obeying the conditions (\cite{TauYu}, p.~410)
\begin{enumerate}
\item $\mathfrak a$ is abelian and all its elements are semisimple,
i.e.\ $\mathfrak a\subset\mathfrak b_s$;
\item as a vector space one has $\mathfrak b=\mathfrak n\oplus\mathfrak a$.
\end{enumerate}

\vspace{12pt}\noindent
{\bf Theorem:} Let $\mathfrak b$ be algebraic and solvable in $\gL(V)$ and let
$\mathfrak n$ be the set of nilpotent endomorphisms of $V$ in $\mathfrak b$.
If $\mathfrak h$ is the maximal commutative Lie subalgebra of $\mathfrak b$
consisting of semisimple elements, $\mathfrak b$ is the semidirect product of
$\mathfrak h$ with $\mathfrak n$ (\cite{TauYu}, p.~455),
\[\mathfrak b=\mathfrak n\rtimes\mathfrak h.\]
The existence of Borel algebras follows from the trianguar decomposition of
the semisimple Lie algebra $\mathfrak g$~\cite{Borel,TauYu,GoodmanWallach,%
Serre:1965,Jacobson:1979,Bauerle:1990,Barut:1986dd},
\[\mathfrak g=\mathfrak n_+\oplus\mathfrak h\oplus\mathfrak n_-\]
where
\[\mathfrak n_+=\sum_{\alpha\in\phi^+}\mathfrak g_\alpha,\qquad
  \mathfrak n_-=\sum_{\alpha\in\phi^+}\mathfrak g_{-\alpha}\]
and $\mathfrak h$ being the Cartan subalgebra of $\mathfrak g$. Then
$\mathfrak b_\pm=\mathfrak h\oplus\mathfrak n_\pm$ are maximal solvable Lie
subalgebras of $\mathfrak g$, called the Borel subalgebra of $\mathfrak g$
relative to $\mathfrak h$. Moreover,
\[\mathfrak n_{\mathfrak g}(\mathfrak b_\pm)=\mathfrak b_\pm\quad\mbox{and}
  \quad\mathfrak z_{\mathfrak g}(\mathfrak b_\pm)=\{0\}\]
where $\mathfrak n_{\mathfrak g}$ is the normaliser and
$\mathfrak z_{\mathfrak g}$ is the centraliser of $\mathfrak b_\pm$ in
$\mathfrak g$.

\subsection{Generators of the Borel subgroup}
It can be easily seen that the exponential operation provides the
parametrisation of the generic element
$\Beps(\vec\xi;\lambda,\omega)\in\Bor_{1,3}$,
\begin{eqnarray}\label{eq19}
\Beps_{e^\lambda}&=&\exp(\tilde\lambda\beps_0)
  \ =\ \1_4+\sinh\tilde\lambda\beps_0+(\cosh\tilde\lambda-1)(\beps_0)^2\qquad
  (\lambda=e^{\tilde\lambda}),\nonumber\\
\Teps_\xi&=&\exp(\vec\xi\vec\beps)\ =\ \1_4+\vec\xi\vec\beps
  +\frac12(\vec\xi\vec\beps)^2\ =\nonumber\\
  &=&\1_4+\xi_1\beps_1+\xi_2\beps_2+\frac12\xi_1^2(\beps_1)^2
  +\frac12\xi_2^2(\beps_2)^2,\nonumber\\
R_\omega&=&\exp(\omega\beps_3)\ =\ \1_4+\sin\omega\beps_3
  +(1-\cos\theta)(\beps_3)^2,
\end{eqnarray}
The general element of $\bor_{1,3}^{(\eps)}$ can be written as
\begin{equation}
Y=y^\mu b_\mu=\pmatrix{0&\eps y^1&\eps y^2&\eps y^0\cr
  \eps y^1&0&-y^3&-y^1\cr \eps y^2&y^3&0&-y^2\cr \eps y^0&y^1&y^2&0\cr}.
\end{equation}
The adjoint representation $\ad Y$ in the basis $\{b_1,b_2,b_0,b_3\}$ of the
semidirect sum $\rtimes$ is calculated to be
\begin{equation}
\ad Y(=\mathop{\rm Reg}Y)=\pmatrix{y^0&-y^3&-y^1&y^2\cr
  y^3&y^0&-y^2&-y^1\cr 0&0&0&0\cr 0&0&0&0\cr}.
\end{equation}
From the secular equation
$\det(\ad Y-\lambda)=(-\lambda)^2((y^0-\lambda)^2+(y^3)^2)$ it follows that
\begin{equation}
\mathop{\rm spec}(\ad Y)=\{0,0,\lambda_3=y^0-iy^3,\lambda_4=y^0+iy^3\}.
\end{equation}
The eigenvalue problem yields two eigenfunctions $Z_3$ and $Z_4$,
\begin{eqnarray}
\ad Y(Z_3)=\lambda_3Z_3&\Rightarrow&Z_3=\frac i2(\beps_1+i\beps_2),\nonumber\\
\ad Y(Z_4)=\lambda_4Z_4&\Rightarrow&Z_4=-\frac i2(\beps_1-i\beps_2),
\end{eqnarray}
i.e.\ $Z_3,Z_4\in\mathfrak t_2^{(\eps)}$. In the special case
$Y=y^0\beps_0+y^3\beps_3$ one obtains
\begin{eqnarray}
[y^0b_0+y^3b_3,Z_3]&=&(y^0-iy^3)Z_3,\nonumber\\\
[y^0b_0+y^3b_3,Z_4]&=&-(y^0+iy^3)Z_4.
\end{eqnarray}
Two cases for $Y$ are important, namely $\teps_0=\frac12(\beps_0+i\beps_3)$
and $\ueps_0=\frac12(\beps_0-i\beps_3)$. Combining in pairs with
$\teps_+=Z_3$ and $\ueps_+=Z_4$,
\begin{eqnarray}\label{eq20}
\teps_0=\frac12(\beps_0+i\beps_3),&&\teps_+=\frac i2(\beps_1+i\beps_2),
  \nonumber\\
\ueps_0=\frac12(\beps_0-i\beps_3),&&\ueps_+=-\frac i2(\beps_1-i\beps_2),
\end{eqnarray}
one obtains the commutation relations
\begin{equation}\label{eq21}
[\teps_0,\teps_+]=\teps_+,\qquad
[\ueps_0,\ueps_+]=\ueps_+,\qquad
[\teps_{0,+},\ueps_{0,+}]=0.
\end{equation}
The elements
\begin{eqnarray}\label{eq22}
\teps_0&=&\frac12(\beps_0+i\beps_3)\ =\ \frac12(\eps e_{03}+ie_{21})
  \ =\ -iJ_3^{(\eps)},\\
\teps_+&=&\frac i2(\beps_1+i\beps_2)\ =\ \frac12(i\eps e_{01}-\eps e_{02}
  +ie_{31}-e_{32})\ =\ J_+^{(\eps)}\nonumber
 \end{eqnarray}
generate the minimal solvable algebra $\sol_2^{(\eps)}(t)$ with underlying
vector space given by $\kest_\R\{\teps_0,\teps_+\}$. Similarly,
\begin{eqnarray}\label{eq23}
\ueps_0&=&\frac12(\beps_0-i\beps_3)\ =\ \frac12(\eps e_{03}-ie_{21})
  \ =\ iK_3^{(\eps)},\\
\ueps_+&=&-\frac i2(\beps_1-i\beps_2)\ =\ -\frac12(i\eps e_{01}+\eps e_{02}
  +ie_{31}+e_{32})\ =\ K_-^{(\eps)}\nonumber
\end{eqnarray}
generate the algebra $\sol_2^{(\eps)}(u)$, and since
$[\sol_2^{(\eps)}(t),\sol_2^{(\eps)}(u)]=0$, one obtains the decomposition
\begin{equation}\label{eq24}
\bor_{1,3}^{(\eps)*}=\sol_2^{(\eps)}(t)\boxplus\sol_2^{(\eps)}(u),
\end{equation}
where $\boxplus$ is the Kronecker sum, $A\boxplus B=A\otimes\1+\1\otimes B$. 

\section{Representations}
Every representation of $\lOR_{1,3}$ defines a particular representation of
the subalgebra $\bor_{1,3}$. Of course, not all the representations are of
that kind but those defined by $\lOR_{1,3}$ are of great importance because
the classification of particles is determined by their Lorentz transformation
properties according to Eqs.~(\ref{eq01}) and~(\ref{eq02}). More precisely,
the common eigenvectors of the representation space of the solvable algebra
$\bor_{1,3}$ are the possible helicity states of the particle.

\vspace{12pt}\noindent
{\bf Theorem:} Let $\mathfrak g$ be a solvable algebra and
$\mathfrak g\to\Gamma(\mathfrak g)$ be a representation on a
finite-dimensional vector space $V$. Then (\cite{Bump}, p.~200)
\begin{enumerate}
\item there exists a vector $v\in V$ which is a simultaneous eigenvector for
  all of $\Gamma(\mathfrak g)$,
\item there exists a basis of $V$ with respect to which all elements of
$\Gamma(\mathfrak g)$ are represented by upper triangular matrices.
\end{enumerate}
Notice that the common eigenvector is determined by all the elements of
$\Gamma(\mathfrak g)$, i.e.\ in our case $\mathfrak g=\bor_{1,3}$ there is no
need to assume $\Gamma(\mathfrak t_2^{(\eps)})=0$. In the complex spaces
\[\kest_\C\{e_{(\mu)}:\
  e_{(\mu)}^\rho=\eta^\rho{}_\mu=\delta_{\rho\mu}\}_0^3,\]
the eigenvectors of the solvable algebra $\sol_2^{(\eps)}(t)$ are
\[\ell_0^{(\eps)}=\eps e_{(0)}+e_{(3)}=(\eps,0,0,1)^T,\qquad
  \ell_1=e_{(1)}+ie_{(2)}=(0,1,i,0)^T.\]
Indeed,
\renewcommand{\theequation}{\arabic{equation}{\it i}}
\begin{eqnarray}\label{eq30i}
\teps_0\ell_0^{(\eps)}=\frac12\ell_0^{(\eps)},&&
\teps_+\ell_0^{(\eps)}=0,\nonumber\\
\teps_0\ell_1=\frac12\ell_1,&&
\teps_+\ell_1=0.
\end{eqnarray}
Accordingly, the eigenvectors of $\sol_2(u)$ are
$\ell_0^{(\eps)}$ and $\ell_2=e_{(1)}-ie_{(2)}=(0,1,-i,0)^T$, where
\addtocounter{equation}{-1}
\renewcommand{\theequation}{\arabic{equation}{\it ii}}
\begin{eqnarray}\label{eq30ii}
\ueps_0\ell_0^{(\eps)}=\frac12\ell_0^{(\eps)},&&
\ueps_+\ell_0^{(\eps)}=0,\nonumber\\
\ueps_0\ell_2=\frac12\ell_2,&&
\ueps_+\ell_2=0.
\end{eqnarray}
Since
\begin{eqnarray}
\ueps_0\ell_1=-\frac12\ell_1,&&
\ueps_+\ell_1=-i\ell_0^{(\eps)},\nonumber\\
\teps_0\ell_2=-\frac12\ell_2,&&
\teps_+\ell_2=i\ell_0^{(\eps)}\nonumber
\end{eqnarray}
and $\bor_{1,3}^*=\sol_2(t)\boxplus\sol_2(u)$, the subspace
$\kest_\C\{\ell_0^{(\eps)},\ell_1,\ell_2\}$ is invariant under the action of
$\bor_{1,3}$. However, the vector $\ell_0^{(\eps)}$ is already the defining
vector for $\bor_{1,3}$ (cf.\ Eq.~(\ref{eq11}). Therefore, there are two
helicity states $\ell_1$ and $\ell_2$ relative to $\ell_0^{(\eps)}$. More
precisely, using the two components $\vec D$ and $\vec B$ of the Lorentz
group defined in Appendix~A, the defining Eq.~(\ref{eq11}) yields the
conditions
\addtocounter{equation}{-1}
\renewcommand{\theequation}{\arabic{equation}{\it iii}}
\begin{equation}\label{eq30iii}
D_3\ell_0^{(\eps)}=0,\qquad B_3\ell_0^{(\eps)}=1\ell_0^{(\eps)},
\end{equation}
\renewcommand{\theequation}{\arabic{equation}}
$\ell_1$ is called right-handed with respect to $\ell_0^{(\eps)}$, and
$\ell_2$ is called left-handed with respect to $\ell_0^{(\eps)}$. The value
$1$ in Eq.~(\ref{eq30iii}) may be considered as helicity~$1$ (not spin
because there is no rotation $\SO_3$). The conditions~(\ref{eq30iii}) are
equivalent to
\[\teps_0\ell_0^{(\eps)}=\ueps_0\ell_0^{(\eps)}=\frac12\ell_0^{(\eps)},\qquad
  \teps_+\ell_0^{(\eps)}=\ueps_+\ell_0^{(\eps)}=0.\]

\subsection{The Chevalley basis}
Depending to the sign of $\eps=\pm1$, the elements of $\sol_2(t)$ and
$\sol_2(u)$ can be represented by elements of the fundamental $\sL_2$ (or
Chevalley) basis
\[h=\pmatrix{1&0\cr 0&-1\cr},\qquad
  e=\pmatrix{0&1\cr 0&0\cr},\qquad
  f=\pmatrix{0&0\cr 1&0\cr}.\]
According to the commutator relations, for $\eps=+1$ one can represent
$\teps_0$ and $\ueps_0$ by $\mp\eps ih/2$, $\teps_+$ and $\ueps_+$ by $ie$
and $if$ -- or vice versa for $\eps=-1$ (cf.\ Eqs.~(\ref{eqA37},\ref{eqA38})).
The two solvable algebras $\sol_2(e)=\kest_\R\{h,ie\}$ and
$\sol_2(f)=\kest_\R\{h,if\}$ are representations of the algebras $\sol_2(t)$
and $\sol_2(u)$, respectively,
\begin{eqnarray}
\teps_0=\left(\frac{1+\eps}4h\right)\boxplus\left(\frac{1-\eps}2h\right)
  =-iJ_3^{(\eps)},&&
\teps_+=\left(\frac{1+\eps}2ie\right)\boxplus\left(\frac{1-\eps}2ie\right)
  =J_+^{(\eps)},\nonumber\\
\ueps_0=-\left(\left(\frac{1-\eps}4h\right)\boxplus\left(\frac{1+\eps}4h\right)
  \right)=iK_3^{(\eps)},&&
\ueps_+=\left(\frac{1-\eps}2if\right)\boxplus\left(\frac{1+\eps}2if\right)
  =K_-^{(\eps)}.\nonumber\\
\end{eqnarray}
This can be generalised to arbitrary representations as for instance the
subrepresentations of the representation $(k,l)$ of $\lOR_{1,3}$. The general
representation $(k,l)$ is given by the rule
\begin{eqnarray}\label{eq31}
\pi^{(k,l)}:\ \bor_{1,3}^{(\eps)*}
  &=&\sol_2^{(\eps)}(e)\boxplus\sol_2^{(\eps)}(f)\to\nonumber\\
  &\to&\pi^{(k,l)}(\bor_{1,3}^{(\eps)*})
  =\pi^{(k)}(\sol_2^{(\eps)}(e))\boxplus\pi^{(l)}(\sol_2^{(\eps)}(f)),
\end{eqnarray}
where
\begin{equation}
\sol_2^{(\eps)}(e)=\frac{1+\eps}2\sol_2(e)\boxplus\frac{1-\eps}2\sol_2(e),\quad
\sol_2^{(\eps)}(f)=\frac{1-\eps}2\sol_2(f)\boxplus\frac{1+\eps}2\sol_2(f).
\end{equation}
By virtue of this construction one obtains
\begin{eqnarray}\label{eq32}
\pi^{(k,l)}(\teps_0)&=&-\left(\frac{i(1+\eps)}4\pi^{(k)}(h)\boxplus
  \frac{i(1-\eps)}4\pi^{(l)}(h)\right)\ =\nonumber\\
  &=&\cases{-\frac i2\pi^{(k)}(h)\otimes\1_l&for $\eps=+1$,\cr
  -\1_k\otimes\frac i2\pi^{(l)}(h)&for $\eps=-1$,\cr}\nonumber\\
\pi^{(k,l)}(\teps_+)&=&\frac{i(1+\eps)}2\pi^{(k)}(e)\boxplus
  \frac{i(1-\eps)}2\pi^{(l)}(e)\ =\nonumber\\
  &=&\cases{i\pi^{(k)}(e)\otimes\1_l&for $\eps=+1$,\cr
  \1_k\otimes i\pi^{(l)}(e)&for $\eps=-1$,\cr}\nonumber\\
\pi^{(k,l)}(\ueps_0)&=&\frac{i(1-\eps)}4\pi^{(k)}(h)\boxplus
  \frac{i(1+\eps)}4\pi^{(l)}(h)\ =\nonumber\\
  &=&\cases{\1_k\otimes\frac i2\pi^{(l)}(h)&for $\eps=+1$,\cr
  \frac i2\pi^{(k)}(h)\otimes\1_l&for $\eps=-1$,\cr}\nonumber\\
\pi^{(k,l)}(\ueps_+)&=&\frac{i(1-\eps)}2\pi^{(k)}(f)\boxplus
  \frac{i(1+\eps)}2\pi^{(l)}(f)\ =\nonumber\\
  &=&\cases{\1_k\otimes i\pi^{(l)}(f)&for $\eps=+1$,\cr
  i\pi^{(k)}(f)\otimes\1_l&for $\eps=-1$,\cr}
\end{eqnarray}
where
\begin{eqnarray}\label{eq33}
\pi^{(k)}(h)|k,m\rangle&=&2m|k,m\rangle,\nonumber\\
\pi^{(k)}(e)|k,m\rangle&=&\rho^{(k)}_{(m)}|k,m+1\rangle,\nonumber\\
\pi^{(k)}(f)|k,m\rangle&=&\rho^{(k)}_{(-m)}|k,m-1\rangle,
\end{eqnarray}
where $\rho^{(k)}_{(m)}=\sqrt{(k-m)(k+m+1)}$. In the representation of
$\pi^{(k,l)}$ by direct products,
\[\{|k,l;m_k,m_l\rangle=|k,m_k\rangle\otimes|l,m_l\rangle:\
  -k\le m_k\le k;\ -l\le m_l\le l\}\]
one obtains common eigenvectors for $\sol_2(t)$ given by
\begin{equation}\label{eq34}
\pi^{(k,l)}(t_0^{(+)})|k,l;k,m_l\rangle=k|k,l;k,m_l\rangle,\qquad
\pi^{(k,l)}(t_+^{(+)})|k,l;k,m_l\rangle=0
\end{equation}
with $m_l=-l,-l+1,\ldots,l$, and common eigenvectors for $\sol_2(u)$
given by
\begin{equation}\label{eq35}
\pi^{(k,l)}(u_0^{(+)})|k,l;m_k,-l\rangle=l|k,l;m_k,-l\rangle,\qquad
\pi^{(k,l)}(u_+^{(+)})|k,l;m_k,-l\rangle=0
\end{equation}
with $m_k=-k,-k+1,\ldots,k$.

\subsection{Resolution of the solvable group}
$\sol_2(e)$ is a Lie algebra of orientation-conserving affine translations
${\rm Aff}_1$. The underlying topological space $\R\times\R_+$ for the
corresponding Lie group $\Sol_2(e)$ is simply connected and open in the
plane $\R^2$. As a general element of this group can be represented by
\[S(\beta,\alpha)=\pmatrix{e^\alpha&\beta\cr 0&1\cr}\]
($\alpha,\beta\in\R$), the geometrical space on which this group acts is the
real line,
\[\Sol_2(e)\ni S(\beta,\alpha):\ \R^1\ni x\to e^\alpha x+\beta\in\R^1.\]
Because of
\[S(\beta,\alpha)S(\beta',\alpha')
  =\pmatrix{e^{\alpha+\alpha'}&e^\alpha\beta'+\beta\cr 0&1\cr}
  =S(e^\alpha\beta'+\beta,\alpha+\alpha'),\]
the solvable group is a semidirect product of two abelian groups,
\[\Sol_2(e)=\R\rtimes\R_+.\]
Therefore, $\Sol_2$ is
\begin{enumerate}
\item locally compact
\item simply connected
\item minimal non-abelian
\item non-compact
\item non-semisimple, i.e.\ solvable,
\item non-unimodular.
\end{enumerate}
Via the exponential mapping, these two parts are generated by $e$ and
$h_+=\frac12(\1+h)$. Therefore, one can write $\sol_2(e)=\R e\rtimes\R h_+$
A similar consideration leads to $\sol_2(f)=\R f\rtimes\R h_-$ where
$h_-=\frac12(\1-h)$. Moreover, $\sol_2(e)$ and $\sol_2(f)$ are related to the
Cartan involution. Therefore, we end up with the amusing two-fold
decomposition
\[\bor_{1,3}^*=(\R e\rtimes\R h_+)\boxplus(\R f\rtimes\R h_-)\]
where
\[h_+=\pmatrix{1&0\cr 0&0\cr},\quad
  e=\pmatrix{0&1\cr 0&0\cr},\quad
  f=\pmatrix{0&0\cr 1&0\cr},\quad
  h_-=\pmatrix{0&0\cr 0&1\cr}.\]
It is important to note that the Kronecker sum structure is imposed by
semisimplicity because $\bor_{1,3}$ is a real form of $\so_4(\C)$, while the
semidirect product structure is caused by solvability.

\subsection{Weinberg's ansatz}
From Steven Weinberg we adopt the following statements which we subsume under
the keyword of ``Weinberg's Ansatz''~\cite{Ohnuki:1976,Weinberg:1995}:
\begin{enumerate}
\item If a massless particle is equal to it's antiparticle, it is described by
the irreducible representation $(k,k)$ of the proper Lorentz group.
\item If a massless particle is not equal to it's antiparticle, the particle
is described by the irreducible representation $(k,0)$ of the proper Lorentz
group, the antiparticle is described by the irreducible representation $(0,k)$
of the proper Lorentz group.
\end{enumerate}
Note that the massless particle is defined via the Borel subgroup by the
irreducible representation of the proper Lorentz group without necessity to
introduce parity separately.

For the representation $(k,k)$, from Eq.~(\ref{eq34}) one obtains the helicity
states associated with $\sol_2^{(+)}(e)$,
\[|k,k;k,-k+p\rangle,\qquad p=0,1,2,\ldots,2k,\]
and from Eq.~(\ref{eq35}) one obtains the helicity states associated with
$\sol_2^{(+)}(f)$,
\[|k,k;k-p,-k\rangle,\qquad p=0,1,2,\ldots,2k.\]
However, the condition~(\ref{eq30iii}) excludes the state $|k,k;k,-k\rangle$,
since
\begin{eqnarray}
D_3^{(k,k)}|k,k;k,-k\rangle&=&i(k-k)|k,k;k,-k\rangle=0,\nonumber\\
B_3^{(k,k)}|k,k;k,-k\rangle&=&2k|k,k;k,-k\rangle.\nonumber
\end{eqnarray}
Therefore, for the particle ($\eps=+1$) with zero mass and helicity
$\lambda=2k$ one obtains the $4k$ helicity states
\begin{equation}\label{eq36}
|k,k;k,-k+p\rangle,\qquad|k,k;k-p,-k\rangle,\qquad p=1,2,\ldots,2k
\end{equation}
relative to the central state $|k,k;k,-k\rangle$ which is determined by the
conditions
\begin{eqnarray}
t_0^{(+)}|k,k;k,-k\rangle&=&u_0^{(+)}|k,k;k,-k\rangle\ =\ k|k,k;k,-k\rangle,
  \nonumber\\
t_+^{(+)}|k,k;k,-k\rangle&=&u_+^{(+)}|k,k;k,-k\rangle\ =\ 0.
\end{eqnarray}
On setting $\eps=-1$ in Eq.~(\ref{eq32}), the helicity states become
\[|k,k;-k+p,k\rangle,\qquad|k,k,-k,k-p\rangle,\qquad p=1,2,\ldots,2k\]
The central state $|k,k;-k,k\rangle$ is defined by the conditions
\begin{eqnarray}
t_0^{(-)}|k,k;-k,k\rangle&=&u_0^{(-)}|k,k;-k,k\rangle\ =\ k|k,k;-k,k\rangle,
  \nonumber\\
t_+^{(-)}|k,k;-k,k\rangle&=&u_+^{(-)}|k,k;-k,k\rangle\ =\ 0.
\end{eqnarray}
Notice that for the vector case $(\frac12,\frac12)$ and $\eps=+1$ the helicity
states are given by
\[\textstyle|\frac12,\frac12;\frac12,\frac12\rangle,\quad
  |\frac12,\frac12;-\frac12,-\frac12\rangle\]
relative to the central state $|\frac12,\frac12;\frac12,-\frac12\rangle$. For
$\eps=-1$ the helicity states are the same, but now relative to the central
state $|\frac12,\frac12;-\frac12,\frac12\rangle$. Therefore, the two
polarisations of the photon as basic quantities in physics are determined by
the proper Lorentz group $\Lor_{1,3}$ without taking refuge to the parity.

As a further example, the massless particle of helicity $\lambda=2$ is
associated with the representation $(1,1)$ and has $4$ helicity states
relative to the state $|1,1;1,-1\rangle$: two right-handed states
$|1,1;1,1\rangle,|1,1;1,0\rangle$ and two left-handed states
$|1,1;0,-1\rangle,|1,1;-1,-1\rangle$.

\subsection{Weyl equations}
The Weyl equations are of the type $(\frac12,0)$ and $(0,\frac12)$. The
representation $(\frac12,0)$ is defined by the commutative
diagram~\cite{Maggiore:2005,Sterman:1994,Yndurain:1996}
\begin{center}
\begin{tabular}{rlcl}
$\Lor_{1,3}\ni\Lambda:$&$\E_{1,3}\ni p^\mu$&$\longrightarrow$
  &$(\Lambda p)^\mu=\Lambda^\mu{}_\nu p^\nu$\\
$\downarrow$&$\downarrow$\rlap{$\sigma$}&&$\downarrow$\rlap{$\sigma$}\\
$\SL_2(\C)\ni\pm A_\Lambda:$&$\H_2\in\sigma(p)=\sigma_\mu p^\mu$
  &$\longrightarrow$&$A_\Lambda^{\phantom\dagger}\sigma(p)A_\Lambda^\dagger$
\end{tabular}\end{center}
where the commutativity of the diagram results in
\[A_\Lambda^{\phantom\dagger}\sigma(p)A_\Lambda^\dagger
  =\sigma(\Lambda p)=\sigma_\mu\Lambda^\mu{}_\nu p^\nu.\]
Using $\tilde\sigma=(\1;-\vec\sigma)$, one obtains
\begin{equation}
A_\Lambda=\frac{\Lambda^{\mu\nu}\sigma_\mu\tilde\sigma_\nu}{2
  \tr(A_\Lambda^\dagger)},\qquad
\Lambda_{A\nu}^\mu=\frac12\tr(\sigma_\mu A_\Lambda\sigma_\nu
  A_\Lambda^\dagger).
\end{equation}
Using the exponential representations
$\Lambda=\exp\left(-\frac12\omega^{\mu\nu}e_{\mu\nu}\right)$ and
$A_\Lambda=\exp\left(-\frac12\omega^{\mu\nu}m_{\mu\nu}\right)$, one obtains a
relation between the generators,
\[m_{\mu\nu}=\frac14(e_{\mu\nu})^{\alpha\beta}\sigma_\alpha\tilde\sigma_\beta
  =\cases{m_{kl}=\frac i2\epsilon_{kl}{}^j\sigma_j\cr
  m_{0j}=\frac12\sigma_j\cr}\]
One can obtain the transformation rule of the Weyl spinor by looking at the
commutative diagram
\begin{center}
\begin{tabular}{rlcl}
$\psi_R:$&$\E_{1,3}\ni x$&$\longrightarrow$&$\psi_R(x)$\\
\llap{$U(\Lambda)$}$\downarrow$&$\downarrow$\rlap{$\Lambda$}&
  &$\downarrow$\rlap{$A_\Lambda$}\\
$U(\Lambda)\psi_R:$&$\E_{1,3}\ni\Lambda x$&$\longrightarrow$
  &$(U(\Lambda)\psi_R)(\Lambda x)$\\
\end{tabular}\end{center}
implying $(U(\Lambda)\psi_R)(\Lambda x)=T^{(1/2,0)}\psi_R(x)
  =A_\Lambda\psi_R(x)$. The Weyl equation
\[\tilde\sigma_\mu\partial^\mu\psi_R(x)=0\]
is given in momentum space by $\tilde\sigma_\mu p^\mu\psi_R(p)=0$. For the
standard vector $\po=(\eps,0,0,1)^T$ this equation reduces to
\begin{equation}\label{eq38}
\sigma_3\ell_R^{(\eps)}=\eps\ell_R^{(\eps)},
\end{equation}
having the solutions $\ell_R^{(\eps)}=(1+\eps)a\ell_1+(1-\eps)b\ell_2$ with
$\ell_1=(1,0)^T$ and $\ell_2=(0,1)^T$. The Borel algebra
$\bor_{1,3}^{(\eps)}(\frac12,0)$ can be expressed as
\begin{eqnarray}
\beps_0({\textstyle\frac12,0})=\frac\eps2h,&&
\beps_1({\textstyle\frac12,0})=\frac{1+\eps}2e-\frac{1-\eps}2f,\nonumber\\
\beps_3({\textstyle\frac12,0})=-\frac i2h,&&
\beps_2({\textstyle\frac12,0})=-\frac{i(1+\eps)}2e-\frac{i(1-\eps)}2f.
\end{eqnarray}
The algebras $\sol_2^{(\eps)}(e)$ and $\sol_2^{(\eps)}(f)$ have the form
\begin{eqnarray}
\sol_2^{(\eps)}(e)&=&\left\{\teps_0({\textstyle\frac12,0})=\frac{1+\eps}4h,\
  \teps_+({\textstyle\frac12,0})=\frac{i(1+\eps)}2e\right\},\nonumber\\
\sol_2^{(\eps)}(f)&=&\left\{\ueps_0({\textstyle\frac12,0})=-\frac{1-\eps}4h,\
  \ueps_+({\textstyle\frac12,0})=\frac{i(1-\eps)}2f\right\}.\nonumber
\end{eqnarray}
The common eigenvector for $\sol_2^{(\eps)}(e)$ is $\ell_1=(1,0)^T$,
\[\teps_0({\textstyle\frac12,0})\ell_1=\frac{1+\eps}4\ell_1,\qquad
  \teps_+({\textstyle\frac12,0})\ell_1=0,\]
and for $\sol_2^{(\eps)}(f)$ one ontains $\ell_2=(0,1)^T$,
\[\ueps_0({\textstyle\frac12,0})\ell_2=-\frac{1-\eps}4\ell_2,\qquad
  \ueps_+({\textstyle\frac12,0})\ell_2=0.\]
Therefore, the eigenvector of $\bor_{1,3}^{(\eps)}(\frac12,0)$ is exactly
equal to the solution of Weyl's equation, where $\ell_1$ is right handed and
$\ell_2$ is left handed. Notice that in case of the irreducible representation
$(\frac12,0)$ of the proper Lorentz group there exists only one single
solution, i.e.\ one helicity state $\lambda=\frac12$.

More generally, the representation space of the Lorentz representation $(k,0)$
is given by
\[V^{(+)}(k,0)=\kest_\C\{|k,0;m,0\rangle:\ m=-k,-k+1,\ldots,k\}.\]
The action of $\bor_{1,3}^{(+)}$ on $V^{(+)}(k,0)$ can be written as
\begin{eqnarray}
t_0^{(+)}|k,0;m,0\rangle=m|k,0;m,0\rangle,&&
u_0^{(+)}|k,0;m,0\rangle=0,\nonumber\\
t_+^{(+)}|k,0;m,0\rangle=i\rho^{(k)}_{(m)}|k,0;m+1,0\rangle,&&
u_+^{(+)}|k,0;m,0\rangle=0.\nonumber
\end{eqnarray}
Therefore, there exists only a single eigenvector
$|k,0;k,0\rangle\in V^{(+)}(k,0)$ of the Borel algebra
$\bor_{1,3}^{(+)}(k,0)$, i.e.\ one a single helicity state with
\begin{eqnarray}
t_0^{(+)}|k,0;k,0\rangle&=&k|k,0;k,0\rangle,\nonumber\\
t_+^{(+)}|k,0;k,0\rangle&=&0,\nonumber\\
u_0^{(+)}|k,0;k,0\rangle&=&u_+^{(+)}|k,0;k,0\rangle\ =\ 0.\nonumber
\end{eqnarray}
For $\eps=-1$ the single helicity state can be written as $|k,0;-k,0\rangle$
with
\begin{eqnarray}
t_0^{(-)}|k,0;-k,0\rangle&=&t_+^{(-)}|k,0;-k,0\rangle\ =\ 0,\nonumber\\
u_0^{(-)}|k,0;-k,0\rangle&=&k|k,0;-k,0\rangle,\nonumber\\
u_+^{(-)}|k,0;-k,0\rangle&=&0.\nonumber
\end{eqnarray}
In the irreducible case $(0,k)$ (and $\eps=+1$) the representation space reads
\[V^{(+)}(0,k)=\kest_\C\{|0,k;0,m\rangle:\ m=-k,-k+1,\ldots,k\},\]
and the action of $\bor_{1,3}^{(+)}(0,k)$ is given by
\begin{eqnarray}
t_0^{(+)}|0,k;0,m\rangle=0,&&
u_0^{(+)}|0,k;0,m\rangle=-m|0,k;0,m\rangle,\nonumber\\
t_+^{(+)}|0,k;0,m\rangle=0,&& 
u_+^{(+)}|0,k;0,m\rangle=i\rho^{(k)}_{(-m)}|0,k;0,m-1\rangle.\nonumber
\end{eqnarray}
The only eigenvector of $\bor_{1,3}^{(+)}(0,k)$ is $|0,k;0,-k\rangle$ with
\begin{eqnarray}
t_0^{(+)}|0,k;0,-k\rangle&=&t_+^{(+)}|0,k;0,-k\rangle\ =\ 0,\nonumber\\
u_0^{(+)}|0,k;0,-k\rangle&=&k|0,k;0,-k\rangle,\nonumber\\
u_+^{(+)}|0,k;0,-k\rangle&=&0.\nonumber
\end{eqnarray}
For $\eps=-1$ one obtains the action
\begin{eqnarray}
t_0^{(-)}|0,k;0,m\rangle=m|0,k;0,m\rangle,&&
u_0^{(-)}|0,k;0,m\rangle=0,\nonumber\\
t_+^{(-)}|0,k;0,m\rangle=i\rho^{(k)}_{(m)}|0,k;0,m+1\rangle,&&
u_+^{(-)}|0,k;0,m\rangle=0\nonumber
\end{eqnarray}
and the only eigenvector $|0,k;0,k\rangle$.

\section{Conclusions}
Turning back to the title of our paper, ``no'' in ``nophysics'' can also stand
for
\begin{itemize}
\item no semisimple (instead, solvable) solution,
\item no abelian (instead, the minimal non-abelian) solution,
\item no compact (but locally compact) solution,
\item no unimodular (instead, non-unimodular) solution,
\item no Killing form (because the Cartan--Killing metric tensor\\
  is identically zero on the derived algebra), and
\item no Casimir invariant found.
\end{itemize}
If semisimplicity, as we are being used to it, stands for ``yes'', solvability
represents ``no''. However, in our opinion, physics as it should be treated is
semisimple as well as solvable, providing a good perspective for our view at
the phenomenon of mass.

Taking solvability as the internal symmetry of massless particles, the photon
of helicity $\lambda=1$ is represented by $(\frac12,\frac12)$, and Pauli's
neutrino and antineutrino of helicity $\lambda=\frac12$ by $(\frac12,0)$ and
$(0,\frac12)$, respectively. Finally, we might ask about the graviton. If the
graviton is identical to its own antiparticle, according to Weinberg's ansatz
it is represented by $(1,1)$ with helicity $\lambda=2$. $(1,0)$ and $(0,1)$
can stand for an hypothetical massless vector boson and its antiboson, both
with helicity $\lambda=1$ but with e.g.\ opposite charge (e.g.\ massless
charges $W$ bosons). However, since open charges for massless particles have
not been seen in any of the accellerator experiments, the validity of
Weinberg's ansatz challenges the graviton as independent particle, i.e.\ the
quantisation of gravity as a whole.

Finally, accepting the Borel algebra as symmetry algebra for massless
particles, it is reasonable to develop a Yang--Mills theory for solvable
groups, treating the elementary algebras $\su_2$ and $\sol_2$ on the same
footing. Indeed, while the algebra $\su_2$ generates semisimple algebras via
the Cartan matrix, the solvable algebras are constructed gradually as
semidirect sums of abelian algebras. Moreover, the solvable gauge will more
immediately generate the abelian gauge of the field theory of electromagnetism,
according to ideas presented by Helfer, Nuyts and others~\cite{Helfer:1987jn,%
Nuyts:2002td,Anco:2015}.

\subsection*{Acknowledgements}
This work was supported by the Estonian Research Council under Grant
No.~IUT2-27. The authors are grateful to Z.~Oziewicz for interesting
discussions on the Lorentz group.

\newpage

\begin{appendix}

\section{The Lorentz group}
\setcounter{equation}{0}\def\theequation{A\arabic{equation}}
The Lorentz group is usually given by its action on the Minkowski vector space
$\E_{1,3}$ with the metric $\eta=\diag(1;-1,-1,-1)$. By definition, the
Lorentz group $O_{1,3}$ preserves the invariant
$x\cdot y=x^\mu\eta_{\mu\nu}y^\nu=x^T\eta y$, $x,y\in\E_{1,3}$ (cf.\ e.g.\
Refs.~\cite{Gelfand:1963,Lyubarskii:1960,Naimark:1964,Sexl:2001,Tung:1999,%
Das:1993,Aste:2016gse}), i.e.\
\begin{equation}\label{eqA01}
\Lambda x\cdot\Lambda y=x^T\Lambda^T\eta\Lambda y\buildrel!\over=x^T\eta y
  =x\cdot y\quad\Rightarrow\quad\Lambda^T\eta\Lambda=\eta,
\end{equation}
where $O_{1,3}\ni\Lambda:\ \E_{1,3}\ni x\to\Lambda x\in\E_{1,3}$. In the
matrix form one obtains
\[(\Lambda x)^\mu=\Lambda^\mu{}_\nu x^\nu\quad\mbox{and}\quad
\Lambda^\mu{}_\nu\eta_{\mu\rho}\Lambda^\rho{}_\sigma=\eta_{\nu\sigma}.\]
The group ${\rm O}_{1,3}$ is topologically homeomorphic to
${\rm O}_3\times\R^3$, and the number of connected components is four.
Henceforth we will consider mainly the component connected to unity,
$\Lor_{1,3}=\SO_{1,3}^0$, called the proper orthochronous Lorentz group,
\begin{equation}\label{eqA02}
\Lor_{1,3}=\{\Lambda\in\M_4(\R):\ \Lambda^T\eta\Lambda=\eta,\
  \det\Lambda=1,\ \Lambda^0{}_0\ge 1\}.
\end{equation}
$\Lor_{1,3}$ is a normal subgroup of ${\rm O}_{1,3}$. The Lorentz group of all
proper orthochronous Lorentz transformations of coordinates on the Minkowski
space is a six-parameter matrix Lie group. The domain of the six parameters is
given by
\[D=\{\eta_1,\eta_2,\eta_3,\omega_1,\omega_2,\omega_3:\
  \eta_i\in\R,\ -\pi<\omega_1\le\pi,\ 0<\omega_2\le\pi,
  -\pi<\omega_3\le\pi\},\]
where the boundary points are topologically identified. The resulting region
is homeomorphic to $\R^3\times\P_3$ where $\P_3$ is the three-dimensional
projective space, covered twice by the simply connected three-dimensional unit
sphere. Therefore, $\Lor_{1,3}$ is locally compact and doubly connected, path
connected, simple and reductive. The universal covering group is $\SL_2(\C)$,
i.e.\
\begin{equation}\label{eqA03}
\Lor_{1,3}=\SL_2(\C)\times\SL_2(\C)/\Z_2=\SO_3(\C)
\end{equation}
where realification of complex groups is understood.

\subsection{Matrix representation}
Assuming the Minkowski metric, it is convenient to write the defining
representation for $\Lambda$ blockwise,
\begin{equation}\label{eqA04}
\Lambda=\pmatrix{A&\vec B^T\cr\vec C&D\cr}
\end{equation}
where $A=\Lambda^0{}_0\ge 1$, $B_k=\Lambda^0{}_k$, $C_k=\Lambda^k{}_0$, and
\[D=(D^i{}_j)\in\GL(3,\R).\]
For $\Lambda\in\Lor_{1,3}$, one has
\renewcommand{\theequation}{A\arabic{equation}.1}
\begin{eqnarray}\label{eqA05.1}
\Lambda^T\eta\Lambda=\eta&\Rightarrow
  &\cases{A^2=1+\vec C^T\vec C=1+|\vec C|^2\cr
  D^TD=\1_3+\vec B\vec B^T\cr A\vec B-D^T\vec C=0\cr}
\end{eqnarray}
\addtocounter{equation}{-1}
\renewcommand{\theequation}{A\arabic{equation}.2}
\begin{eqnarray}\label{eqA05.2}
\Lambda\eta\Lambda^T=\eta&\Rightarrow
  &\cases{A^2=1+\vec B^T\vec B=1+|\vec B|^2\cr
  DD^T=\1_3+\vec C\vec C^T\cr A\vec C-D\vec B=0}
\end{eqnarray}
\renewcommand{\theequation}{A\arabic{equation}}
Using these equations, it is easy to see that
\[\Lambda^{-1}=\pmatrix{A&-\vec C^T\cr -\vec B&D^T\cr}.\]
Because of $\det\Lambda=1$, one has $\det D=A\ge 1$. One can use the equations
to rewrite
\[\vec C=\frac1AD\vec B\quad\mbox{or}\quad\vec B=\frac1AD^T\vec C\]
to obtain
\begin{equation}\label{eqA06}
\Lambda=\pmatrix{A&\vec B^T\cr \frac1AD\vec B&D\cr}
  =\pmatrix{A&\frac1A\vec C^TD\cr \vec C&D\cr}.
\end{equation}
As a consequence of this, one can apply a polar
decomposition~\cite{Serre:2002,Ungar:1991pca,Ungar:2015} to the elements of
the Lorentz group $\Lor_{1,3}$, $\Lambda=QP=P'Q$ where $Q$ is orthogonal and
$P,P'$ are real symmetric positive definite. Using the ansatz
\begin{equation}\label{eqA07}
Q=\pmatrix{1&0\cr 0&R\cr},\quad P=\pmatrix{A&\vec B^T\cr \vec B&D_P\cr},\quad
P'=\pmatrix{A&\vec C^T\cr\vec C&D'_P\cr}
\end{equation}
with $R\in\SO_3$ and $D_P,D'_P$ symmetric, one obtains
\begin{equation}\label{eqA08}
R=\frac1{1+A}(D+AD^{-1T}),\quad
D_P=\frac1{1+A}(A+D^TD),\quad
D'_P=\frac1{1+A}(A+DD^T).
\end{equation}
According to {\em Tolhoek's theorem\/}~\cite{Jauch:1959dbc},
$P=(\Lambda^T\Lambda)^{1/2}$ and $P'=(\Lambda\Lambda^T)^{1/2}$
describe pure Lorentz transformations or boosts, where
\[A=\frac1{\sqrt{1-v^2/c^2}},\qquad\vec C=\frac1cA\vec v.\]
Vice versa, $\Lambda=QP(\vec B)=P(R\vec B)Q$ can be written as
\begin{equation}\label{eqA10}
\Lambda=\pmatrix{A&\vec B^T\cr R\vec B&R+\frac1{1+A}R\vec B\vec B^T\cr}
  =\pmatrix{A&\vec C^TR\cr\vec C&R+\frac1{1+A}\vec C\vec C^TR\cr},
\end{equation}
where $A=\Lambda^0{}_0\ge 1$, $\vec B,\vec C\in\R^3$ and $R\in\SO_3$.
The {\em Principal axis theorem\/} for the group $\Lor_{1,3}$, finally, tells
us that every $\Lambda\in\Lor_{1,3}$ has one of the shapes
\begin{eqnarray}
S\Lambda_sS^{-1}&=&S\pmatrix{\lOR t&0\cr 0&\rot\omega\cr}S^{-1}\quad\mbox{or}
  \nonumber\\
S\Lambda_uS^{-1}&=&S\pmatrix{1&0\cr 0&N\cr}S^{-1},
\end{eqnarray}
where
\begin{eqnarray}\label{eqA11}
\lOR t&=&\pmatrix{\cosh t&\sinh t\cr\sinh t&\cosh t\cr},\quad t\in\R,
  \nonumber\\
\rot\omega&=&\pmatrix{\cos\omega&-\sin\omega\cr\sin\omega&\cos\omega\cr}
  \in\SO_2,\quad -\pi<\omega\le\pi.
\end{eqnarray}
$S\in\Lor_{1,3}$ is a similarity transformation. Since
$\spec\Lambda_s=\{e^t,e^{-t},e^{i\omega},e^{-i\omega}\}$, $\Lambda_s$ is
semisimple. On the contrary $\Lambda_u$ us unipotent. Therefore, the Jordan
form of $N$ is
\[\pmatrix{1&1&0\cr 0&1&1\cr 0&0&1\cr}\]
and $\spec\Lambda_u=\{1\}$. Finally, we note that for all
$\Lambda\in\Lor_{1,3}$, $\Lambda$ satisfies the minimal
equation~\cite{Hanson:2011xx}
\[\Lambda^4-(\tr\Lambda)\Lambda^3+\frac12\left((\tr\Lambda)^2-\tr\Lambda^2
  \right)\Lambda^2-(\tr\Lambda)\Lambda+\1_4=0.\]

Though it is a pure algebraic reason, one concludes from polar decomposition
that the maximal compact subgroup of $\Lor_{1,3}$ is isomorphic to $\SO_3$.
Indeed, $\Lor_{1,3}$ is isomorphic to $\SO_3(\C)$, and $\SO_3$ is the compact
real form of the latter. Moreover, since $\Lor_{1,3}$ is stable under the
transposition (i.e.\ $\Lambda\in\Lor_{1,3}\Rightarrow\Lambda^T\in\Lor_{1,3}$),
$\Lor_{1,3}$ is a  linear reductive group, for which the maximal compact
subgroup $K$ is determined by the Cartan involution
\[\theta:\ \Lor_{1,3}\ni\Lambda\to\theta(\Lambda)=\Lambda^{-1T}\in\Lor_{1,3}\]
as $K=\{\Lambda\in\Lor_{1,3}:\ \theta(\Lambda)=\Lambda^{-1T}=\Lambda\}=\SO_3$.
In this setting, $\Lor_{1,3}/\SO_3$ is a symmetric space.

It is important that the maximal simple compact subgroup
$\SO_3\subset\Lor_{1,3}$ determines the internal symmetry of massive
particles, i.e.\ the spin. On the other hand, the maximal solvable
noncompact subgroup called Borel subgroup $\Bor_{1,3}\subset\Lor_{1,3}$
determines the helicity of massless particles. Notice that $\Bor_{1,3}$ is a
semidirect product of the abelian subgroups ${\cal T}_2$ and $\Tor_{1,3}$,
$\Bor_{1,3}={\cal T}_2\rtimes\Tor_{1,3}$, where $\Tor_{1,3}$ is the maximal
Torus of $\Lor_{1,3}$.

\subsection{Generators of $\Lor_{1,3}$\label{lor13gen}}
In order to linearise the group $\Lor_{1,3}$, one can simply differentiate it
and evaluate the derivative at the identity element of the group. The tangent
space at the identity element is the Lie algebra
\begin{equation}\label{eqA12}
\lOR_{1,3}=\{X\in\M_4(\R):\ e^{tX}\in\Lor_{1,3}\forall t\in\R\}.
\end{equation}
According to {\em Lie's theorem\/}, the exponential map
$\exp:\ \lOR_{1,3}\to\Lor_{1,3}$ is surjective. Therefore, any element
$\Lambda\in\Lor_{1,3}$ that is close to unity can be written as the
exponential of an element $X\in\lOR_{1,3}$. From Eq.~(\ref{eqA01}) one
concludes that the defining equation for an element $X\in\lOR_{1,3}$ is given
by
\begin{equation}\label{eqA13}
X^T\eta+\eta X=0.
\end{equation}
Using the infinitesimal transformation
\begin{equation}\label{eqA14}
\Lambda^\mu{}_\nu=\eta^\mu{}_\nu+\omega^\mu{}_\nu
  \equiv\eta^\mu{}_\nu-\frac12(\omega_{\rho\sigma}e^{\rho\sigma})^\mu{}_\nu,
\end{equation}
the defining equation~(\ref{eqA13}) gives $\omega_{\mu\nu}=-\omega_{\nu\mu}$,
and one obtains six independent parameters $\omega_{\mu\nu}$ and generators
$e^{\mu\nu}=-e^{\nu\mu}$. A generic element $\Lambda\in\Lor_{1,3}$ is written
as
\begin{equation}\label{eqA15}
\Lambda(\omega)=\exp\left(-\frac12\omega^{\mu\nu}e_{\mu\nu}\right),\qquad
e^{\mu\nu}=-\frac\partial{\partial\omega^{\mu\nu}}\Lambda(\omega)
\Big|_{\omega=0}.
\end{equation}
The six independent generators $e^{\mu\nu}$ have the form
\begin{equation}\label{eqA16}
(e^{\mu\nu})^\rho{}_\sigma=-\eta^{\mu\rho}\eta^\nu{}_\sigma
  +\eta^{\nu\rho}\eta^\mu{}_\sigma
\end{equation}
and obey the commutation relation
\begin{equation}\label{eqA17}
[e^{\mu\nu},e^{\rho\sigma}]=\eta^{\mu\rho}e^{\nu\sigma}
  +\eta^{\nu\sigma}e^{\mu\rho}-\eta^{\mu\sigma}e^{\nu\rho}
  -\eta^{\nu\rho}e^{\mu\sigma}.
\end{equation}
The defining equation~(\ref{eqA13}) applies to the generators in the form
\begin{equation}\label{eqA18}
e^{\mu\nu T}\eta+\eta e^{\mu\nu}=0.
\end{equation}
The minimal equation for $X\in\lOR_{1,3}(\equiv\so_{1,3})$ is given by
\[X^4-\frac12(\tr X^2)X^2+(\det X)\1_4=0,\]
and $\det X\le 0$, $\tr X=0$.

\subsection{Cartan decomposition}
Following Ref.~\cite{Gilmore:2002}, let $\vec e_{(i)}$ ($i=1,2,3$) be an
orthogonal triad for $\R^3$ defined by\footnote{The notation
$(X^T)_\mu\equiv X^\mu$ is used in the following.}
\[(\vec e_{(i)}^T)_j=(\vec e_{(i)})^j=\delta_i^j,\qquad
\vec e_{(i)}^T\vec e_{(j)}^{\phantom T}=\delta_{ij},\qquad
(\vec e_{(i)}\times\vec e_{(j)})^k=\epsilon_{ij}{}^k\]
($1=\epsilon^{0123}=-\epsilon_{0123}\equiv-\epsilon_{123}$). In terms of this
triad the generators $e^{\mu\nu}$ are given by
\[e_{ij}=\pmatrix{0&0\cr 0&f_{ij}\cr}\equiv\epsilon_{ijk}D^k\]
or, vice versa,
\renewcommand{\theequation}{A\arabic{equation}.1}
\begin{equation}\label{eqA19.1}
D_i=-\frac12\epsilon_{0ijk}e^{jk}=\pmatrix{0&0\cr 0&-\epsilon_i{}_{jk}
  \vec e_{(j)}^{\phantom T}\vec e_{(k)}^T\cr}=-D^i,
\end{equation}
where $(D_i)^\mu{}_\nu=\epsilon_{0j}{}^\mu{}_\nu$,
$f_{ij}=\vec e_{(i)}^{\phantom T}\vec e_{(j)}^T
  =\vec e_{(j)}^{\phantom T}\vec e_{(i)}^T$, and
\addtocounter{equation}{-1}
\renewcommand{\theequation}{A\arabic{equation}.2}
\begin{equation}\label{eqA19.2}
e_{0i}=\pmatrix{0&\vec e_{(i)}^T\cr\vec e_{(i)}&0\cr}\equiv B_i,
\end{equation}
\renewcommand{\theequation}{A\arabic{equation}}
where $(B_i)^\mu{}_\nu=-\eta_0{}^\mu\eta_{i\nu}+\eta_{0\nu}\eta_i{}^\mu$. A
general element $X\in\lOR_{1,3}$ has the form
\begin{equation}\label{eqA20}
X=\vec\omega\vec D-\vec\eta\vec B=\pmatrix{0&\eta^1&\eta^2&\eta^3\cr
  \eta^1&0&-\omega^3&\omega^2\cr \eta^2&\omega^3&0&-\omega^1\cr
  \eta^3&-\omega^2&\omega^1&0\cr}.
\end{equation}
The corresponding finite Lorentz transformation is given by
\begin{equation}\label{eqA21}
\Lambda=\exp\left(-\frac12\omega^{\mu\nu}e_{\mu\nu}\right)
  =\exp(\vec\omega\vec D-\vec\eta\vec B),
\end{equation}
where $\omega^i=\frac12\epsilon^{ijk}\omega_{jk}$,
$\eta^i=\omega_{0i}=-\eta_i$. The commutation relations can be expressed as
\begin{eqnarray}\label{eqA22}
[D_i,D_j]&=&\epsilon_{ijk}D_k,\nonumber\\\
[D_i,B_j]&=&\epsilon_{ijk}B_k,\nonumber\\\
[B_i,B_j]&=&-\epsilon_{ijk}D_k.
\end{eqnarray}
The compact generators $D_i$ are antisymmetric ($D_i^T=-D_i$) while the
noncompact generators $B_i$ are symmetric ($B_i^T=B_i$). As a consequence, the
Lorentz algebra $\lOR_{1,3}$ (if considered as vector space) is a symmetric
Lie algebra with symmetric decomposition 
\begin{equation}\label{eqA24}
\vec{\mathstrut\lOR}_{1,3}=\vec{\mathstrut\so}_3\oplus
  \vec{\mathstrut\mathfrak p},
\end{equation}
where $\vec{\mathstrut\mathfrak p}=\kest_\R\{B_i\}_1^3$. Indeed, $\so_3$ is a
subalgebra, $[\so_3,\so_3]=\so_3$, but $[\so_3,\mathfrak p]\subset\mathfrak p$
and $[\mathfrak p,\mathfrak p]\subset\so_3$. Given the structure~(\ref{eqA20}),
the generic element $X\in\lOR_{1,3}$ can be split up into two parts,
\begin{equation}
X=\pmatrix{0&\vec X^T\cr\vec X&X_{(3)}\cr}
  =\pmatrix{0&0\cr 0&X_{(3)}\cr}+\pmatrix{0&\vec X^T\cr\vec X&0\cr}
\end{equation}
where the first part is compact, $X_{(3)}^T=-X^{(3)}$, and contained in
$\so_3$, the second part is noncompact and contained in $\mathfrak p$. Notice
that $\so_3$ and $\mathfrak p$ are orthogonal with respect to the Killing form,
\begin{equation}
(\so_3,\mathfrak p)=0.
\end{equation}
Since $\lOR_{1,3}$ is simple, the Cartan--Killing form is nonsingular on
$\so_3$ and $\mathfrak p$. The symmetric decomposition is determined by the
Cartan involution
\begin{equation}\label{eqA26}
\theta:\ \lOR_{1,3}\ni X\to\theta(X)=-X^T=\eta X\eta\in\lOR_{1,3}
\end{equation}
which is the only external involtive automorphism for $\lOR_{1,3}$. Indeed,
one obtains $\theta(D_i)=-D_i^T=D_i$ and, therefore,
\renewcommand{\theequation}{A\arabic{equation}.1}
\begin{equation}\label{eqA27.1}
\so_3=\{X\in\lOR_{1,3}:\ \theta(X)=+X\}=\kest_\R\{D_i\}_1^3
\end{equation}
is the maximal compact subalgebra of $\lOR_{1,3}$. Similarly,
\addtocounter{equation}{-1}
\renewcommand{\theequation}{A\arabic{equation}.2}
\begin{equation}
\mathfrak p=\{X\in\lOR_{1,3}:\ \theta(X)=-X\}=\kest_\R\{B_i\}_1^3
\end{equation}
\renewcommand{\theequation}{A\arabic{equation}}
consists of the noncompact elements of $\lOR_{1,3}$. Therefore, one ends up
with the Cartan decomposition
\begin{equation}\label{eqA28}
\lOR_{1,3}=\so_3+\mathfrak p.
\end{equation}

The map $\SO_3\times\mathfrak p\to\Lor_{1,3}$ given by
\[SO_3\times\mathfrak p\ni(R,X)\to R\exp X\in\Lor_{1,3}\]
is a diffeomorphism onto $\Lor_{1,3}$. Therefore, the Cartan
decomposition~(\ref{eqA28}) on the level of the Lie algebra $\lOR_{1,3}$
induces the polar decomposition~(\ref{eqA07}) on the level of the Lie group
$\Lor_{1,3}$. The exponential map is a diffeomorphism from the vector space
$\vec{\mathfrak p}$ of symmetric matrices to the set of positive definite
matrices,
\[\exp:\ \vec{\mathfrak p}\ni X=X^T\to\exp X=(\exp X)^T.\]

\subsection{Weyl's unitary trick}
It is an algebraic fact that the symmetric spaces appear in pairs. If the
Cartan involution induces $\mathfrak g=\mathfrak k+\mathfrak p$, it's
companion is
\begin{equation}\label{eqA29}
\mathfrak g^{(W)}=\mathfrak k+i\mathfrak p\equiv\mathfrak g_\C
  =\C\otimes\mathfrak g.
\end{equation}
If $\mathfrak g$ is the Lie algebra of a noncompact connected semisimple Lie
group $G$, $\mathfrak g^{(W)}$ is the Lie algebra of a second Lie group
$G^{(W)}$ which is compact. In this way the noncompact algebras appearing in
the Cartan decomposition can be analytically continued to compact algebras by
analytic extension,
\begin{equation}
\lOR_{1,3}=\so_3+\mathfrak p\to\so_3+i\mathfrak p\equiv\lOR_{1,3}^{(W)}.
\end{equation}
This analytical continuation known as Weyl's unitary trick can be accomplished
by using the matrix
\[\Gamma=\pmatrix{i&0\cr 0&\1_3\cr}\]
in the way
\[\lOR_{1,3}\ni X\buildrel\Gamma\over\to X^{(W)}=\Gamma X\Gamma\]
(note that Weyl's unitary trick is not a similarity transformation). One
obtains
\begin{equation}\label{eqA30}
X^{(W)}=\Gamma X\Gamma=\pmatrix{0&i\vec X^T\cr i\vec X&X_{(3)}\cr}
  =\Gamma^{-1}(-\eta X)\Gamma.
\end{equation}
and for the basis
\begin{equation}
D_i^{(W)}=\Gamma D_i\Gamma=D_i,\qquad B_i^{(W)}=\Gamma B_i\Gamma=iB_i.
\end{equation}
Accordingly, the commutation relations~(\ref{eqA22}) change to their compact
form
\begin{eqnarray}\label{eqA32}
[D_i^{(W)},D_j^{(W)}]&=&\epsilon_{ijk}D_k^{(W)},\nonumber\\\
[D_i^{(W)},B_j^{(W)}]&=&\epsilon_{ijk}B_k^{(W)},\nonumber\\\
[B_i^{(W)},B_j^{(W)}]&=&\epsilon_{ijk}D_k^{(W)}.
\end{eqnarray}
Because of $(\eta X)^T=-\eta X$, one recovers the algebra $\so_4(\R)$. The
last algebra in turn is isomorphic to $\su_2\boxplus\su_2$ where $\boxplus$
denotes the Kronecker sum of algebras, i.e.\ for $a\in\mathfrak g$ and
$b\in\mathfrak h$ one has
\begin{equation}\label{eqA35}
a\boxplus b\equiv a\otimes\1_{\mathfrak h}+\1_{\mathfrak g}\otimes b
  \in\mathfrak g\boxplus\mathfrak h.
\end{equation}
To conclude, the pair of symmetric algebras $\lOR_{1,3}$ and $\lOR_{1,3}^{(W)}$
is connected by Weyl's unitary trick,
\begin{equation}\label{eqA34}
\lOR_{1,3}\buildrel{\rm Weyl}\over\longrightarrow\lOR_{1,3}^{(W)}
  =S(\su_2\boxplus\su_2)S^\dagger.
\end{equation}
The splitting map
\begin{equation}\label{eqA36}
S=\frac1{\sqrt2}\pmatrix{0&1&-1&0\cr -1&0&0&1\cr -i&0&0&-i\cr 0&1&1&0\cr}
\end{equation}
with $S^\dagger=S^{-1}$ gives the decomposition in the $\su_2$ basis,
\begin{eqnarray}\label{eqA37}
D_i^{(W)}\to S^\dagger D_i^{(W)}S=m_i\boxplus m_i&\Rightarrow&
  D_i=m_i\boxplus m_i,\nonumber\\[7pt]
B_i^{(W)}\to S^\dagger B_i^{(W)}S=m_i\boxplus(-m_i)&\Rightarrow&
  B_i=(-im_i)\boxplus(im_i),
\end{eqnarray}
where $m_j=\frac i2\sigma_j$, and the decomposition in the $\sL_2$ basis,
\begin{eqnarray}\label{eqA38}
D_1=\frac i2(e\boxplus e)+\frac i2(f\boxplus f),&&
B_1=\frac12\left(e\boxplus(-e)\right)+\frac12\left(f\boxplus(-f)\right),
  \nonumber\\
D_2=\frac12(e\boxplus e)-\frac12(f\boxplus f),&&
B_2=-\frac i2\left(e\boxplus(-e)\right)+\frac i2\left(f\boxplus(-f)\right),
  \nonumber\\
D_3=\frac i2(h\boxplus h),&&
B_3=\frac12\left(h\boxplus(-h)\right).
\end{eqnarray}
The meaning of Weyl's unitary trick is that the representations of the
noncompact group $\Lor_{1,3}$ may be viewed as representations of the compact
group $\SO_4=\SU_2\times\SU_2/\Z_2$, where
$\Z_2=\{(\1_2,\1_2),(-\1_2,-\1_2)\}$ is the discrete subgroup. From the
unitary nature of the representations of the compact group $\SO(4)$ one
concludes the full reducibility of the finite-dimensional representations of
$\Lor_{1,3}$. Note that $\SO_4$ is the real compact form of
$\SO_4(\C)=\SL_2(\C)\times\SL_2(C)/\Z_2$, and $\Lor_{1,3}$ is a real noncompact
form. Since the simply connected group $\SU(2)\times\SU(2)$ is a universal
covering group of the doubly connected group $\SO_4$, their Lie algebras are
isomorphic. Moreover, since $\su_2$ is a compact real form of $\sL_2(\C)$,
the construction of the representations of the algebra $\lOR_{1,3}$ may be
realised by using the representations of $\sL_2(\C)$. Since $\SL_2(\C)$ as
the topological product $\R^3\times\SU(2)$ is simply connected, all its
representations are single-valued.

\subsection{Higher dimensional representations}
In the standard basis
\begin{equation}\label{eqA39}
h=\pmatrix{1&0\cr 0&-1\cr},\qquad
e=\pmatrix{0&1\cr 0&0\cr},\qquad
f=\pmatrix{0&0\cr 1&0\cr}
\end{equation}
of $\sL_2(\C)$ with $[h,e]=2e$, $[h,f]=-2f$ and $[e,f]=h$, the
$(2k+1)$-dimensional, real representations applied to states $|k,m\rangle$ are
given by
\begin{eqnarray}\label{eqA41}
\pi^{(k)}(h)|k,m\rangle&=&2m|k,m\rangle,\nonumber\\
\pi^{(k)}(e)|k,m\rangle&=&\rho^{(k)}_{(m)}|k,m+1\rangle,\nonumber\\
\pi^{(k)}(f)|k,m\rangle&=&\rho^{(k)}_{(-m)}|k,m-1\rangle,
\end{eqnarray}
where $k=0,\frac12,1,\ldots$, $m=-k,-k+1,\ldots,k$ and
$\rho^{(k)}_{(m)}=\sqrt{(k-m+1)(k+m)}$.

\vspace{12pt}\noindent
{\bf Theorem:} Let $2k\in\N$ and let $(V,\pi)$ be a simple representation of
$\sL_2(\C)$ of dimension $2k+1$. Then (\cite{TauYu}, p.~281)
\begin{enumerate}
\item $\pi$ is equivalent to $\pi^{(k)}$,
\item the eigenvalues of $\pi^{(k)}(h)/2$ are
  $\{-k,-k+1,\ldots,k\}=\spec\frac12\pi^{(k)}(h)$,
\item if $0\ne v\in V$ verifies $\pi^{(k)}(e)v=0$, then $\pi^{(k)}(h)v=2kv$,\\
  i.e.\ $\pi^{(k)}(h)$ and $\pi^{(k)}(e)$ have the common eigenvector
  $|k,k\rangle$,
\item if $0\ne v\in V$ verifies $\pi^{(k)}(f)v=0$, then $\pi^{(k)}(h)v=-2kv$,\\
  i.e.\ $\pi^{(k)}(h)$ and $\pi^{(k)}(f)$ have the common eigenvector
  $|k,-k\rangle$
\end{enumerate}
Since $\su(2)$ is the compact real form of $\sL_2(C)$ and the generators of
$\su(2)$ are given by
\begin{equation}\label{eqA42}
m_1=\frac i2(e+f),\qquad m_2=\frac12(e-f),\qquad m_3=\frac i2h,
\end{equation}
one can accordingly define irreducible representations of $\su(2)$ given by
\begin{equation}\label{eqA43}
\pi^{(k)}(m_1),\quad\pi^{(k)}(m_2)\quad\mbox{and}\quad
\pi^{(k)}(m_3).
\end{equation}
Following the procedure given before, the real Lie algebra $\lOR_{1,3}$ may
be identified via Weyl's unitary trick with the algebra $\lOR_{1,3}^{(W)}$
which splits into a Kronecker sum of two algebras $\su(2)$,
$\lOR_{1,3}^{(W)}\sim\su_2\boxplus\su_2$. If $\pi^{(k)}$ and $\pi^{(l)}$ are
representations of $\su_2$ on the vector spaces $V^{(k)}$ and $V^{(l)}$,
$\pi^{(k)}\otimes\pi^{(l)}$ is a representation of the Lie algebra
$\lOR_{1,3}^{(W)}$ on $V^{(k)}\otimes V^{(l)}$, defined by
\begin{equation}\label{eqA44}
m_i\boxplus m_j\to\pi^{(k,l)}(m_i\boxplus m_j)
  =\pi^{(k)}(m_i)\boxplus\pi^{(l)}(m_j).
\end{equation}
The representation $\pi^{(k,l)}$ of the Kronecker sum $\su_2\boxplus\su_2$ on
the tensor product basis
\[\{|k,l;m_k,m_l\rangle\equiv|k,m_k\rangle\otimes|l,m_l\rangle,
  -k\le m_k\le k,\ -l\le m_l\le l\}\]
is given by
\begin{eqnarray}\label{eqA45}
\lefteqn{\pi^{(k,l)}(m_i\boxplus m_j)|k,l;m_k,m_l\rangle\ =}\nonumber\\
  &=&\sum_{m=-k}^k\left(\pi^{(k)}(m_i)\right)_{mm_k}|k,l;m,m_l\rangle
  +\sum_{m=-l}^l\left(\pi^{(l)}(m_j)\right)_{mm_l}|k,l;m_k,m\rangle.
\end{eqnarray}
Using the representations of $D_i,B_i\in\lOR_{1,3}$ in the $\su_2$ basis, one
obtains
\begin{eqnarray}\label{eqA46}
\pi^{(k,l)}(D_i)&=&\pi^{(k)}(m_i)\boxplus\pi^{(l)}(m_i),\nonumber\\
\pi^{(k,l)}(B_i)&=&\left(-i\pi^{(k)}(m_i)\right)
  \boxplus\left(i\pi^{(l)}(m_i)\right)
\end{eqnarray}
with $\pi^{(k)}(m_i)$ given by Eq.~(\ref{eqA43}).

\subsection{Splitting algebra}
{\bf Theorem:} Any finite-dimensional representation of $\lOR_{1,3}$ is
isomorphic to $\pi^{(k,l)}$ for some $k,l=0,\frac12,1,\ldots$ and is
non-antihermitean. The corresponding representation of $\Lor_{1,3}$ is
non-unitary~\cite{TauYu}.

\vspace{12pt}
The isomorphism between $\so_4$ and $\so_3\oplus\so_3$ is easily realised
by the choice of the basis
\begin{eqnarray}\label{eqA47}
J_i^{(\eps)}&=&\frac12(D_i+i\eps B_i)
  \ =\ \frac{1+\eps}2m_i\boxplus\frac{1-\eps}2m_i,\nonumber\\
K_i^{(\eps)}&=&\frac12(D_i-i\eps B_i)
  \ =\ \frac{1-\eps}2m_i\boxplus\frac{1+\eps}2m_i
\end{eqnarray}
with $J_i^{(\eps)\dagger}=-J_i^{(\eps)}$, $K_i^{(\eps)\dagger}=-K_i^{(\eps)}$
($\eps=\pm 1$). The commutator relations are given by
\begin{eqnarray}\label{eqA48}
[J_i^{(\eps)},J_j^{(\eps)}]&=&\epsilon_{ijk}J_k^{(\eps)},\nonumber\\\
[J_i^{(\eps)},K_j^{(\eps)}]&=&0,\nonumber\\\
[K_i^{(\eps)},K_j^{(\eps)}]&=&\epsilon_{ijk}K_j^{(\eps)}.
\end{eqnarray}
Note that the fact that the Lorentz algebra $\lOR_{1,3}$ can be written as a
Kronecker sum $\su(2)\boxplus\su(2)$ of two algebras does not mean that
$\lOR_{1,3}$ is the same as $\su(2)\oplus\su(2)$ or $\lOR_{1,3}^{(W)}$.
Rather, they are the anti-hermitean complex representations of $\lOR_{1,3}$.

\subsection{Spinor representations}
There are two fundamental spinor representations, from which all other may be
obtained by tensor product reduction~\cite{Joos:1962qq,Corson:1953,%
Macfarlane:1962,Ticciati:1999}. The Lorentz covariant description needs two
sets of relativistic Pauli matrices,
\[(\sigma_\mu)=(\1_2,\vec\sigma)\quad\mbox{and}\quad
  (\tilde\sigma_\mu)=(\1_2,-\vec\sigma).\]
The relation between the real Minkowski space $\E_{1,3}$ and the set of all
complex hermitean $2\times 2$ matrices $\H_2$ is given by
\begin{equation}\label{eqA48p}
\E_{1,3}\ni p\to\sigma(p)=\sigma_\mu p^\mu\in\H_2.
\end{equation}
The correspondence is a linear isomorphism,
\begin{equation}\label{eqA49}
\det\sigma(p)=p^2=p^\mu p_\mu,
\end{equation}
and the characteristic polynomial
\begin{equation}\label{eqA50}
\det(\sigma(p)-\lambda)=\cases{(p^0+|\vec p\,|-\lambda)(p^0-|\vec p\,|-\lambda)
  &for $p^2>0$,\cr \lambda(2p^2-\lambda)&for $p^2=0$.\cr}
\end{equation}
Therefore, if $p^2>0$, $\sigma(p)$ is positive semidefinite.

\vspace{12pt}\noindent
{\bf Theorem:}\begin{enumerate}
\item Let $\sigma(p)\in\M_2(\C)$ be positive definite.\\ If $A\in\M_n(\C)$ and
$\det A\ne 0$, then $A\sigma(p)A^\dagger$ is positive definite~\cite{TauYu}.
\item If $\sigma(p)\in\M_2(\C)$ is not positive definite but positive
semidefinite and if $A\in\M_n(\C)$, then $A\sigma(p)A^\dagger$ is always
positive semidefinite and not positive definite~\cite{TauYu}.
\end{enumerate}
The fundamental representation $(\frac12,0)$ can be expressed as the
commutative diagram
\begin{equation}\label{eqA51}
\begin{tabular}{rlcl}
$\Lor_{1,3}\ni\Lambda:$&$\E_{1,3}\ni p$&$\longrightarrow$&$\Lambda p$\\
$\downarrow$&$\downarrow$\rlap{$\sigma$}&&$\downarrow$\rlap{$\sigma$}\\
$\SL_2(\C)\ni\pm A_\Lambda:$&$\H_2\ni\sigma(p)$&$\longrightarrow$
  &$A_\Lambda^{\phantom\dagger}\sigma(p)A_\Lambda^\dagger=\sigma(\Lambda p)$
\end{tabular}
\end{equation}
The continuous homomorphism relates an element $\Lambda\in\Lor_{1,3}$ to two
elements $\pm A_\Lambda\in\SL_2(\C)$. The group $\SL_2(\C)$ constitutes the
universal covering group of $\Lor_{1,3}$, i.e.\ $\Lor_{1,3}=\SL_2(\C)/\Z$
(on the right hand side the realification of $\SL_2(\C)$ is understood). Using
Pauli's spin matrices one obtains
\begin{eqnarray}\label{eqA52}
A_\Lambda^{\phantom\dagger}\sigma_\nu A_\Lambda^\dagger
  &=&\sigma_\mu\Lambda^\mu{}_\nu,\nonumber\\
\Lambda^\mu{}_\nu&=&\frac12\tr(\sigma_\mu A_\Lambda^{\phantom\dagger}
  \sigma_\nu A_\Lambda^\dagger),\nonumber\\
A_\Lambda&=&\frac{\Lambda^{\mu\nu}\sigma_\mu\tilde\sigma_\nu}{(\frac12
  \Lambda^{\alpha\beta}\Lambda^{\gamma\delta}\tr(\sigma_\alpha
  \tilde\sigma_\beta\sigma_\delta\tilde\sigma_\gamma))^{1/2}},
\end{eqnarray}
where
\[4(\tr A_\Lambda^\dagger)^2=\frac12\Lambda^{\mu\nu}\Lambda^{\rho\sigma}
  \tr\left(\sigma_\mu\tilde\sigma_\nu\sigma_\sigma\tilde\sigma_\rho\right)
  =(\tr\Lambda)^2-\tr\Lambda^2+4+i\Lambda^{\mu\nu}\Lambda^{\rho\sigma}
  \epsilon_{\mu\nu\rho\sigma}\]
(note the different order of indices). Suppose that $\Lambda$ and $A_\Lambda$
are given by $A_\Lambda^{\phantom\dagger}\sigma_\nu A_\Lambda^\dagger
=\sigma_\mu\Lambda^\mu{}_\nu$, one can write
\begin{eqnarray}
A_\Lambda^\dagger\sigma_\nu A_\Lambda^{\phantom\dagger}
  &=&\sigma_\mu(\Lambda^T)^\mu{}_\nu,\nonumber\\
A_\Lambda^\dagger\tilde\sigma_\nu A_\Lambda^{\phantom\dagger}
  &=&\tilde\sigma_\mu(\Lambda^{-1})^\mu{}_\nu,\nonumber\\
A_\Lambda^{\phantom\dagger}\tilde\sigma_\nu A_\Lambda^\dagger
  &=&\tilde\sigma_\mu(\Lambda^{-1T})^\mu{}_\nu.\nonumber
\end{eqnarray}
Defining $\Lambda=\exp(-\frac12\omega^{\mu\nu}e_{\mu\nu})$ and
$A_\Lambda=\exp(-\frac12\omega^{\mu\nu}e_{\mu\nu})$, for the representation
$(\frac12,0)$ one obtains
\begin{eqnarray}\label{eqA53}
m_{\mu\nu}&=&\frac14(e_{\mu\nu})^{\alpha\beta}\sigma_\alpha\tilde\sigma_\beta
  \ =\ -\frac14(\sigma_\mu\tilde\sigma_\nu-\sigma_\nu\tilde\sigma_\mu),
  \nonumber\\
m_i&=&-\frac12\epsilon_{ijk}m^{jk}\ =\ \frac i2\sigma_i,\nonumber\\
m_{0j}&=&\frac12\sigma_j\ =\ -im_j.
\end{eqnarray}
The second fundamental representation $(0,\frac12)$ is defined by the
commutative diagram
\[\begin{tabular}{rlcl}
$\Lor_{1,3}\ni\Lambda:$&$\E_{1,3}\ni p$&$\longrightarrow$&$\Lambda p$\\
$\downarrow$&$\downarrow$\rlap{$\tilde\sigma$}&
  &$\downarrow$\rlap{$\tilde\sigma$}\\
$\SL_2(\C)\ni\pm B_\Lambda:$&$\H_2\ni\tilde\sigma(p)$&$\longrightarrow$
  &$B_\Lambda^{\phantom\dagger}\tilde\sigma(p)B_\Lambda^\dagger
  =\tilde\sigma(\Lambda p)$
\end{tabular}\]
where $\tilde\sigma=C^{-1}\sigma^TC=C^{-1}\sigma^*C$,
\[C=-i\sigma_2=\pmatrix{0&-1\cr 1&0\cr}\]
and $\tilde\sigma(p)=2p_0-\sigma(p)$. The properties of the Pauli matrices
yield
\begin{eqnarray}
B_\Lambda^{\phantom\dagger}\tilde\sigma_\nu B_\Lambda^\dagger
  &=&\tilde\sigma_\mu\Lambda^\mu{}_\nu,\nonumber\\
\Lambda^\mu{}_\nu&=&\frac12\tr(\tilde\sigma_\mu B_\Lambda^{\phantom\dagger}
  \tilde\sigma_\nu B_\Lambda^\dagger),
  \nonumber\\
B_\Lambda&=&\frac{\Lambda^{\mu\nu}\tilde\sigma_\mu\sigma_\nu}{(\frac12
  \Lambda^{\alpha\beta}\Lambda^{\gamma\delta}\tr(\tilde\sigma_\alpha
  \sigma_\beta\tilde\sigma_\delta\sigma_\gamma))^{1/2}}\nonumber
\end{eqnarray}
and
\[4(\tr B_\Lambda^\dagger)^2=\frac12\Lambda^{\mu\nu}\Lambda^{\rho\sigma}
  \tr(\tilde\sigma_\mu\sigma_\nu\tilde\sigma_\sigma\sigma_\rho)
  =(\tr\Lambda)^2-\tr\Lambda^2+4-i\Lambda^{\mu\nu}\Lambda^{\rho\sigma}
  \epsilon_{\mu\nu\rho\sigma}.\]
Defining the generators of the representation $(0,\frac12)$ by
$B_\Lambda=\exp\left(-\frac12\omega^{\mu\nu}\tilde m_{\mu\nu}\right)$,
one obtains
\begin{eqnarray}
\tilde m_{\mu\nu}&=&\frac14(e_{\mu\nu})^{\alpha\beta}\tilde\sigma_\alpha
  \sigma_\beta\ =\ -\frac14(\tilde\sigma_\mu\sigma_\nu
  -\tilde\sigma_\nu\sigma_\mu),\nonumber\\
\tilde m_i&=&-\frac12\epsilon_{ijk}\tilde m^{jk}\ =\ -\frac i2\tilde\sigma_i
  \ =\ \frac i2\sigma_i,\nonumber\\
\tilde m_{0j}&=&\frac12\tilde\sigma_j\ =\ -\frac12\sigma_j\ =\ i\tilde m_j.
  \nonumber
\end{eqnarray}
The two nonequivalent fundamental representations $A_\Lambda$ and $B_\Lambda$
are related by
\[B_\Lambda=C^{-1}A_\Lambda^\dagger C=\left(A_\Lambda\right)^{-1\dagger}.\]

\end{appendix}

\end{document}